\documentclass[%
  reprint,
  superscriptaddress,
  amsmath,amssymb,
  aps,
  prx,
  longbibliography,
]{revtex4-2}

\usepackage{graphicx}
\usepackage{dcolumn}
\usepackage{bm}
\usepackage{braket}
\usepackage{amsmath}
\usepackage{amssymb}
\usepackage{xcolor}      % before hyperref
\usepackage{tikz}
\usepackage{quantikz}
\usepackage{adjustbox}
\usepackage[colorlinks=true,allcolors=blue]{hyperref}
\usepackage{orcidlink}
\providecommand{\ii}{\mathrm{i}}

\begin{document}

\preprint{APS/123-QED}

\title{End-to-End Molecular Dynamics with a Langevin Thermostat on Quantum Circuits}

\author{Masari Watanabe\orcidlink{0000-0003-3186-535X}}
\email{mwatanabe@quemix.com}
\affiliation{Quemix Inc., Taiyo Life Nihonbashi Building, 2-11-2,
Nihonbashi Chuo-ku, Tokyo 103-0027, Japan}
\affiliation{Department of Physics, The University of Tokyo, Hongo, Bunkyo-ku, Tokyo 113-0033, Japan}

\author{Hirofumi Nishi\orcidlink{0000-0001-5155-6605}}
\affiliation{Quemix Inc., Taiyo Life Nihonbashi Building, 2-11-2,
Nihonbashi Chuo-ku, Tokyo 103-0027, Japan}
\affiliation{Department of Physics, The University of Tokyo, Hongo, Bunkyo-ku, Tokyo 113-0033, Japan}

\author{Taichi Kosugi\orcidlink{0000-0003-3379-3361}}
\affiliation{Quemix Inc., Taiyo Life Nihonbashi Building, 2-11-2,
Nihonbashi Chuo-ku, Tokyo 103-0027, Japan}
\affiliation{Department of Physics, The University of Tokyo, Hongo, Bunkyo-ku, Tokyo 113-0033, Japan}

\author{Shigekazu Hidaka\orcidlink{0000-0002-7992-1018}}
\affiliation{Advanced Research and Innovation Center, DENSO CORPORATION, 500-1 Minamiyama, Komenoki-cho, Nisshin, Aichi 470-0111, Japan}

\author{Ryo Sakurai\orcidlink{0009-0004-2721-7231}}
\affiliation{Advanced Research and Innovation Center, DENSO CORPORATION, 1-1-4, Haneda Airport, Ota-ku, Tokyo 144-0041, Japan}

\author{Yu-ichiro Matsushita\orcidlink{0000-0002-9254-5918}}
\affiliation{Quemix Inc., Taiyo Life Nihonbashi Building, 2-11-2, Nihonbashi Chuo-ku, Tokyo 103-0027, Japan}
\affiliation{Department of Physics, The University of Tokyo, Hongo, Bunkyo-ku, Tokyo 113-0033, Japan}
\affiliation{Quantum Materials and Applications Research Center,
National Institutes for Quantum Science and Technology, Tokyo 152-8550, Japan}

\date{\today}

\begin{abstract}
We construct a quantum-circuit framework for finite-temperature molecular
dynamics in the canonical ensemble (NVT) with a Langevin thermostat, connecting
canonical state preparation to subsequent physical-property readouts. The classical nuclear phase-space
distribution is encoded as a Koopman--von Neumann (KvN) wave function, and
canonical state preparation is formulated as Langevin-type Fokker--Planck
relaxation. The Hamiltonian Liouville flow, momentum friction, and momentum
diffusion are decomposed into separate circuit blocks. The friction block is
represented by a symmetrized momentum-space dilation, whereas the diffusion
block is implemented as a cosine filter realized by probabilistic imaginary-time
evolution (PITE). We analytically quantify the leading-order temperature bias
caused by replacing the Gaussian diffusion kernel with this PITE-realized
cosine filter. This analysis yields an internal-temperature correction that
targets the desired physical equilibrium distribution.
As a proof-of-concept demonstration connecting quantum chemistry to KvN nuclear
dynamics, we study the H$_2$ molecule. Numerical simulations show relaxation
from a nonequilibrium phase-space distribution to a canonical KvN state. From
this canonical state, we demonstrate two complementary readouts: a dynamical
quantum-phase-estimation readout of the vibrational density of states associated
with the H--H stretch coordinate and a static canonical evaluation of the
transition-state-theory (TST) rate constant. This work demonstrates, in a
minimal molecular system, a circuit-level protocol that connects Langevin
canonical state preparation to physical-property calculations, providing a
concrete step toward quantum--classical hybrid molecular dynamics on quantum
computers.
\end{abstract}

\maketitle

%==============================================================================
\section{\label{sec:intro}Introduction}
%==============================================================================

Recent progress in quantum hardware and quantum algorithms has accelerated efforts
toward practically useful quantum computing~\cite{GoogleQAI2025Nature,Bluvstein2024Nature,Rodriguez2025Nature,Ransford2025Helios,Montanaro2016npjQI}.
Quantum chemistry is one of the major target applications~\cite{McArdle2020}. Full
configuration interaction (FCI) provides an exact solution within a chosen
one-particle basis, but its computational cost grows rapidly with the number of
electrons and basis functions. Accelerating high-accuracy quantum-chemistry
calculations, including FCI, has therefore become an important frontier in
quantum-algorithm research~\cite{AspuruGuzik2005,McArdle2020}.

Combining Born--Oppenheimer (BO) potential-energy surfaces and forces from
quantum-chemistry calculations with the equations of motion for nuclei yields
\textit{ab initio} molecular dynamics (AIMD)~\cite{Car1985,MarxHutter2009}.
Molecular dynamics (MD) is a foundational method for analyzing chemical reactions,
catalytic processes, materials properties, and biomolecular functions at finite
temperature through nuclear motion~\cite{MarxHutter2009,Tuckerman2010}.
A quantum-computing framework for MD therefore must go beyond
electronic-structure calculations alone, connecting electronic-structure output to
finite-temperature nuclear dynamics and providing readout protocols for MD
observables, such as vibrational spectra and reaction rates. The central goal of
this paper is to construct such a framework, in which potential-energy surfaces and
forces from electronic-structure calculations are used to prepare a canonical
nuclear phase-space distribution on quantum circuits and to read out
molecular-dynamics observables from that distribution.

Existing approaches to MD on quantum computers fall broadly into two categories. In
the first, the BO approximation is retained, the electronic-structure calculation is
evaluated on a quantum computer, and the nuclear motion is propagated on a classical
computer~\cite{Fedorov2021JCP,Dononelli2026}. This route is naturally compatible
with near-term quantum devices, but the nuclear phase-space distribution remains
classical data. In the second, electrons and nuclei are treated on the same quantum
footing, with the goal of evolving the full molecular wave function beyond the BO
approximation~\cite{Kassal2008,daJornada2025FTQC,Pocrnic2026preBO,Eklund2026endtoend}.
This route is powerful in principle for nonadiabatic dynamics, but it currently
presumes large-scale fault-tolerant quantum computers.

The present work takes an intermediate route. We retain the BO approximation and
treat the nuclear degrees of freedom classically, while encoding their phase-space
distribution as a state on quantum registers. The Koopman--von Neumann (KvN)
formalism provides a natural mathematical foundation for this
representation~\cite{Koopman1931,vonNeumann1932,Joseph2020PRR}. In the KvN
formalism, classical Liouville dynamics is represented as linear evolution on a
Hilbert space, so a classical phase-space distribution can be encoded in the
amplitudes of a quantum state. This makes it possible to evolve the phase-space
distribution itself, rather than sampling many classical trajectories and
reconstructing the distribution afterward. In addition, combining a prepared KvN
state with readout schemes such as quantum phase estimation or
amplitude-estimation-type protocols provides a route to direct estimates of
vibrational spectra and observables controlled by low-probability regions of phase
space.

In our previous work, we demonstrated a quantum-circuit evaluation of classical
transport coefficients based on the KvN formalism, while treating canonical state
preparation as an oracle~\cite{PaperI}. For finite-temperature MD, however,
canonical state preparation must be supplied as an explicit circuit primitive. To
address this issue, we focus on Langevin-type Fokker--Planck evolution. A Langevin
thermostat realizes temperature control without introducing Nos\'{e}-type extended
thermostat variables as explicit quantum registers, which helps keep the qubit
requirement modest. Moreover, because the Langevin equation contains both momentum
friction and diffusion, its Fokker--Planck evolution can be used as a
state-preparation mechanism that relaxes a nonequilibrium phase-space distribution
toward a canonical distribution.

\begin{figure*}[htb]
  \centering
    \includegraphics[width=\textwidth]{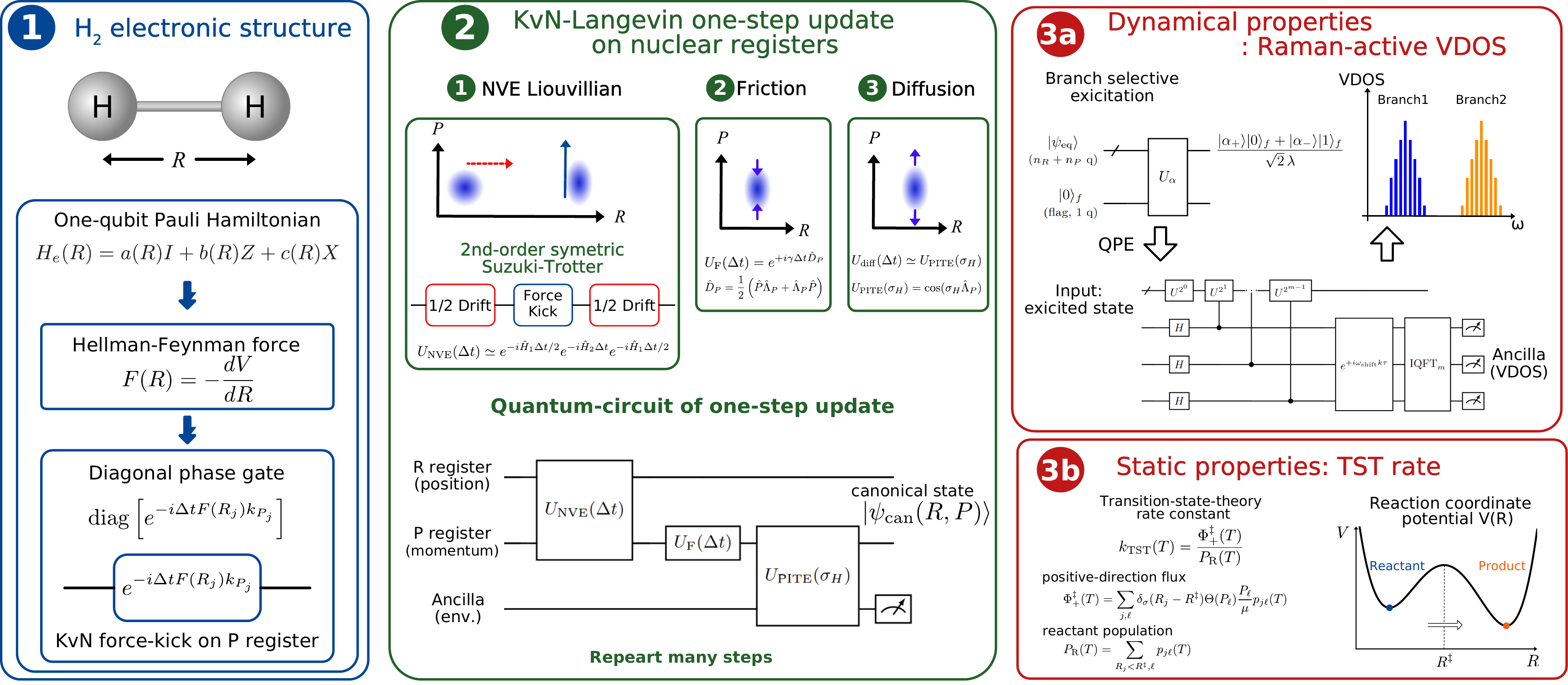}
  \caption{\label{fig:pipeline}%
    \textbf{End-to-end Koopman--von Neumann (KvN)--Langevin molecular-dynamics
    pipeline.} The H$_2$ electronic structure provides the Born--Oppenheimer
    potential and Hellmann--Feynman force, which are embedded as an
    $R$-dependent KvN force kick. Repeated KvN--Langevin updates prepare a
    canonical nuclear phase-space state, from which two observables are read out:
    the Raman-active vibrational density of states by quantum phase estimation and
    the transition-state-theory rate constant from a canonical flux expectation.}
\end{figure*}

In this work, we construct a KvN--Langevin quantum-circuit pipeline that connects
canonical nuclear phase-space state preparation to molecular-dynamics observable
readout. Langevin dynamics is decomposed into Hamiltonian Liouville, momentum
friction, and momentum diffusion blocks. The friction block is represented as a
symmetrized momentum-space dilation acting unitarily on the KvN amplitude, while
the diffusion block is implemented by a cosine filter based on probabilistic
imaginary-time evolution (PITE) and corrected for its leading-order temperature
bias. For H$_2$, the BO potential and Hellmann--Feynman force are obtained from a
one-qubit Pauli Hamiltonian and embedded as an $R$-dependent KvN force-kick
phase. We then demonstrate two readouts from the prepared canonical KvN state: a
quantum-phase-estimation (QPE) readout of the vibrational density of states
(VDOS) associated with the Raman-active H--H stretch and a static canonical
evaluation of the transition-state-theory (TST) rate constant.
Figure~\ref{fig:pipeline} summarizes the overall workflow.

The remainder of this paper is organized as follows. In Sec.~\ref{sec:framework} we
introduce the KvN--Langevin circuit and the PITE/cos-filter implementation of the
diffusion term. In Sec.~\ref{sec:qc-interface} we present the numerical
implementation that supplies the BO potential and Hellmann--Feynman force from the
one-qubit Pauli Hamiltonian of H$_2$ to the KvN force kick. In
Sec.~\ref{sec:canonical} we verify canonical state preparation and bond-formation
dynamics. In Sec.~\ref{sec:vdos} we present the readout of the VDOS associated with
a Raman-active vibrational mode. In Sec.~\ref{sec:tst} we evaluate the TST rate
constant as a static canonical observable. Finally, in
Sec.~\ref{sec:conclusion-outlook} we summarize our conclusions and outlook.

%==============================================================================
\section{\label{sec:framework}KvN--Langevin framework}
%==============================================================================
In this section, we formulate finite-temperature nuclear dynamics under the
Born--Oppenheimer approximation in the Koopman--von Neumann (KvN) representation and
decompose the resulting Langevin evolution into circuit blocks. We consider a
one-dimensional nuclear coordinate $R$ moving on a potential $V(R)$, with conjugate
momentum $P$ and reduced mass $\mu$. We first introduce the minimal KvN structure for
the thermostat-free Hamiltonian flow, then describe the Fokker--Planck splitting of
the Langevin thermostat, and finally give the quantum-circuit implementation of the
corresponding one-step update shown in Fig.~\ref{fig:kvn-langevin-circuit}.

\subsection{\label{sec:kvn-minimal}KvN representation}

In the KvN formalism~\cite{Koopman1931,vonNeumann1932,Joseph2020PRR}, a classical
phase-space density is represented through a complex amplitude $\psi(R,P,t)$ as
\begin{equation}
  \rho(R,P,t) = |\psi(R,P,t)|^{2}.
  \label{eq:kvn-density}
\end{equation}
The phase-space coordinates $R$ and $P$ are mutually commuting operators and are
taken to be diagonal in the computational basis. Their conjugate differential
operators are
\begin{equation}
  \hat{\Lambda}_{R} = -i\partial_{R}, \qquad
  \hat{\Lambda}_{P} = -i\partial_{P}.
  \label{eq:kvn-lambda}
\end{equation}

We begin with the thermostat-free Hamiltonian flow. For the classical Hamiltonian
$H(R,P)=P^{2}/(2\mu)+V(R)$ and force $F(R)=-\partial_{R}V(R)$, the Liouville flow
preserves phase-space volume and therefore induces unitary evolution of the KvN
amplitude. The Hermitian generator of the constant-energy, or NVE, part is
\begin{equation}
  \hat{H}_{\mathrm{NVE}}
  = \frac{\hat{P}}{\mu}\hat{\Lambda}_{R}
  + F(\hat{R})\hat{\Lambda}_{P}
  \equiv \hat{H}_{1} + \hat{H}_{2},
  \label{eq:H-nve}
\end{equation}
with the corresponding propagator
$U_{\mathrm{NVE}}(\Delta t)=
\exp[-i\Delta t\,\hat{H}_{\mathrm{NVE}}]$.
This unitary represents the Hamiltonian Liouville evolution of the nuclear
phase-space amplitude in the absence of a thermostat. Its circuit implementation is
given in Sec.~\ref{sec:quantum-circuit-implementation}.

\subsection{\label{sec:langevin-splitting}Langevin thermostat and Fokker--Planck splitting}

To prepare a finite-temperature canonical distribution, we introduce a Langevin
thermostat. We denote the target physical temperature by $T_{\mathrm{phys}}$, the
friction coefficient by $\gamma$, and set $k_{B}=1$. The Langevin equations are
\begin{align}
  \dot{R} &= \frac{P}{\mu}, \label{eq:langevin-R}\\
  \dot{P} &= -\partial_{R}V(R) - \gamma P
  + \sqrt{2\mu\gamma T_{\mathrm{phys}}}\,\eta(t), \label{eq:langevin-P}
\end{align}
where $\eta(t)$ is a white-noise process satisfying
$\langle\eta(t)\rangle=0$ and
$\langle \eta(t)\eta(t') \rangle = \delta(t-t')$. The associated Fokker--Planck
generator for the density is decomposed as~\cite{Risken1989,LeimkuhlerMatthews}
\begin{align}
  \partial_{t}\rho &= \mathcal{L}_{\mathrm{FP}}\rho, \quad
  \mathcal{L}_{\mathrm{FP}}
  = \mathcal{L}_{\mathrm{NVE}}
  + \mathcal{L}_{\mathrm{fric}}
  + \mathcal{L}_{\mathrm{diff}}, \label{eq:fp-decomposition}\\
  \mathcal{L}_{\mathrm{NVE}}\rho
  &= -\frac{P}{\mu}\partial_{R}\rho
  + \partial_{R}V(R)\,\partial_{P}\rho, \label{eq:L-nve}\\
  \mathcal{L}_{\mathrm{fric}}\rho
  &= \gamma\,\partial_{P}(P\rho), \label{eq:L-fric}\\
  \mathcal{L}_{\mathrm{diff}}\rho
  &= \mu\gamma T_{\mathrm{phys}}\,\partial_{P}^{2}\rho.
  \label{eq:L-diff}
\end{align}
Here $\mathcal{L}_{\mathrm{NVE}}$ generates the Hamiltonian Liouville flow,
$\mathcal{L}_{\mathrm{fric}}$ contracts the momentum distribution, and
$\mathcal{L}_{\mathrm{diff}}$ produces thermal momentum diffusion. Their sum has the
canonical stationary distribution
\[
\rho_{\mathrm{eq}}(R,P) \propto
\exp\!\left[-\frac{P^{2}/2\mu + V(R)}{T_{\mathrm{phys}}}\right],
\]
as required by the fluctuation--dissipation relation.

The NVE part of the Fokker--Planck generator is the density-level counterpart of the
KvN unitary introduced in Sec.~\ref{sec:kvn-minimal}. In other words,
$\partial_{t}\rho=\mathcal{L}_{\mathrm{NVE}}\rho$ is obtained from the amplitude
evolution
$\ket{\psi(t+\Delta t)}=
U_{\mathrm{NVE}}(\Delta t)\ket{\psi(t)}$.

The friction term generates a deterministic contraction of the momentum
coordinate, $P\mapsto e^{-\gamma\Delta t}P$. At the density level, the
corresponding Fokker--Planck substep is governed by
\begin{equation}
\begin{aligned}
\partial_t \rho = \gamma \partial_P(P\rho).
\end{aligned}
\end{equation}
The exact solution over one time step is
\begin{equation}
\begin{aligned}
\rho_{t+\Delta t}(R,P)
=
e^{\gamma\Delta t}
\rho_t\!\left(R,e^{\gamma\Delta t}P\right).
\end{aligned}
\end{equation}
The prefactor is the Jacobian of the inverse momentum map and ensures
conservation of the total probability. Thus, the phase-space flow is
compressible and narrows the momentum distribution. In particular, the second
moment transforms as
\begin{equation}
\begin{aligned}
\langle P^2\rangle_{t+\Delta t}
=
e^{-2\gamma\Delta t}\langle P^2\rangle_t .
\end{aligned}
\end{equation}
In the KvN circuit, this density-level contraction is represented by the
unitary friction block $U_{\mathrm F}$, whose symmetrized amplitude-level
implementation is introduced in
Sec.~\ref{sec:quantum-circuit-implementation}.

The diffusion term describes the stochastic momentum kicks from the Langevin
thermostat. Taken in isolation, it gives the heat equation in the momentum
direction,
\begin{equation}
\begin{aligned}
\partial_t \rho
=
D_P \partial_P^2 \rho,
\qquad
D_P = \mu\gamma T_{\mathrm{phys}},
\end{aligned}
\label{eq:diffusion-substep}
\end{equation}
where $D_P$ is the momentum-space diffusion coefficient. Let $k_P$ be the
Fourier-conjugate variable to $P$, or equivalently the eigenvalue of
$\Lambda_P=-i\partial_P$, and let $\widetilde{\rho}(R,k_P,t)$ be the Fourier
transform of the density along the momentum direction. Over a time step
$\Delta t$, the ideal diffusion substep acts as
\begin{equation}
\begin{aligned}
\widetilde{\rho}(R,k_P,t+\Delta t)
=
\exp\!\left[-D_P\Delta t\,k_P^2\right]
\widetilde{\rho}(R,k_P,t).
\end{aligned}
\label{eq:density-heat-kernel}
\end{equation}
Thus, diffusion damps high-frequency components of the momentum distribution
and smooths the density. Equivalently, this substep increases the momentum
variance according to
\begin{equation}
\begin{aligned}
\langle P^2\rangle_{t+\Delta t}
=
\langle P^2\rangle_t
+
2D_P\Delta t .
\end{aligned}
\label{eq:diffusion-second-moment}
\end{equation}
In the full Langevin dynamics, this increase in momentum variance balances the
contraction caused by friction, yielding the Maxwell--Boltzmann momentum width
at the target temperature in the stationary state.

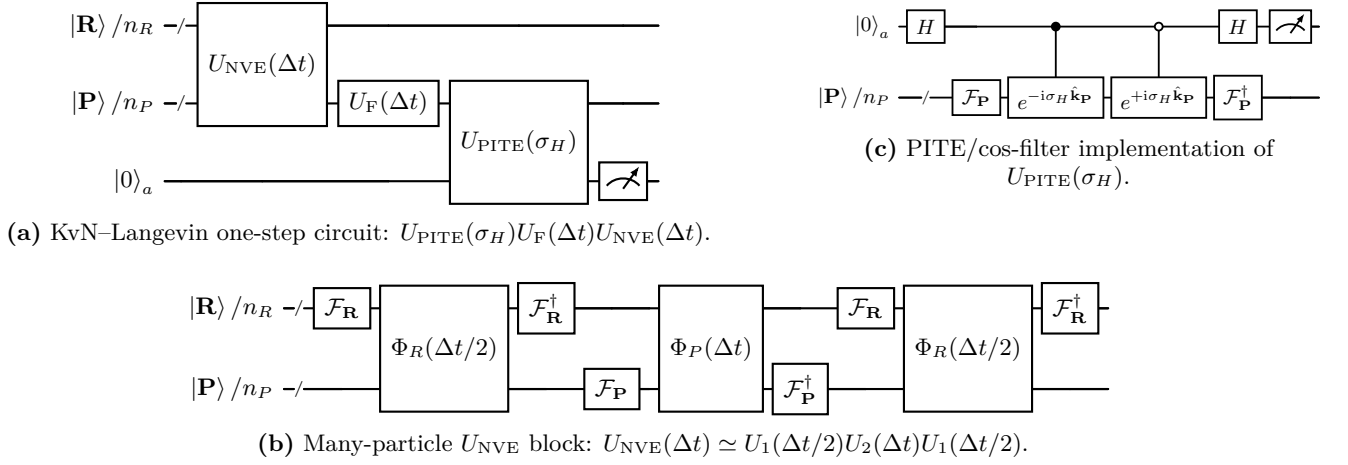
\begin{figure*}[t!h]
  \centering
  \begin{minipage}[t]{0.58\textwidth}
    \vspace{0pt}
    \centering
    \begin{adjustbox}{max width=\linewidth,center}
    \begin{quantikz}[font=\small, column sep=0.15cm, row sep=0.43cm, line width=0.55pt]
      \lstick{$\ket{\mathbf R}/n_R$}
        & \qw \push{\scriptstyle /}
        & \gate[wires=2]{U_{\rm NVE}(\Delta t)}
        & \qw
        & \qw
        & \qw
        & \qw \\
      \lstick{$\ket{\mathbf P}/n_P$}
        & \qw \push{\scriptstyle /}
        & \ghost{U_{\rm NVE}(\Delta t)}
        & \gate{U_{\rm F}(\Delta t)}
        & \gate[wires=2]{U_{\rm PITE}(\sigma_H)}
        & \qw
        & \qw \\
      \lstick{$\ket{0}_a$}
        & \qw
        & \qw
        & \qw
        & \ghost{U_{\rm PITE}(\sigma_H)}
        & \meter{}
        & \qw
    \end{quantikz}
    \end{adjustbox}

    \vspace{0.32em}
    {\small
    \textbf{(a)} KvN--Langevin one-step circuit:
    $U_{\rm PITE}(\sigma_H)U_{\rm F}(\Delta t)U_{\rm NVE}(\Delta t)$.
    }
  \end{minipage}
  \hfill
  \begin{minipage}[t]{0.38\textwidth}
    \vspace{0pt}
    \centering
    \begin{adjustbox}{max width=\linewidth,center}
    \begin{quantikz}[font=\small, column sep=0.12cm, row sep=0.50cm, line width=0.55pt]
      \lstick{$\ket{0}_a$}
        & \gate{H}
        & \qw
        & \ctrl{1}
        & \octrl{1}
        & \gate{H}
        & \meter{}
        & \qw \\
      \lstick{$\ket{\mathbf P}/n_P$}
        & \qw \push{\scriptstyle /}
        & \gate{\mathcal F_{\mathbf P}}
        & \gate{e^{-\ii\sigma_H\hat{\mathbf k}_{\mathbf P}}}
        & \gate{e^{+\ii\sigma_H\hat{\mathbf k}_{\mathbf P}}}
        & \gate{\mathcal F_{\mathbf P}^{\dagger}}
        & \qw
        & \qw
    \end{quantikz}
    \end{adjustbox}

    \vspace{0.36em}
    {\small \textbf{(c)} PITE/cos-filter implementation of $U_{\rm PITE}(\sigma_H)$.}
  \end{minipage}

  \vspace{1.1em}

  \begin{minipage}[t]{0.95\textwidth}
    \vspace{0pt}
    \centering
    \begin{adjustbox}{max width=\linewidth,center}
    \begin{quantikz}[font=\small, column sep=0.13cm, row sep=0.40cm, line width=0.55pt]
      \lstick{$\ket{\mathbf R}/n_R$}
        & \qw \push{\scriptstyle /}
        & \gate{\mathcal F_{\mathbf R}}
        & \gate[wires=2]{\Phi_R(\Delta t/2)}
        & \gate{\mathcal F_{\mathbf R}^{\dagger}}
        & \qw
        & \qw
        & \gate[wires=2]{\Phi_P(\Delta t)}
        & \qw
        & \gate{\mathcal F_{\mathbf R}}
        & \gate[wires=2]{\Phi_R(\Delta t/2)}
        & \gate{\mathcal F_{\mathbf R}^{\dagger}}
        & \qw \\
      \lstick{$\ket{\mathbf P}/n_P$}
        & \qw \push{\scriptstyle /}
        & \qw
        & \ghost{\Phi_R(\Delta t/2)}
        & \qw
        & \gate{\mathcal F_{\mathbf P}}
        & \qw
        & \ghost{\Phi_P(\Delta t)}
        & \gate{\mathcal F_{\mathbf P}^{\dagger}}
        & \qw
        & \ghost{\Phi_R(\Delta t/2)}
        & \qw
        & \qw
    \end{quantikz}
    \end{adjustbox}

    \vspace{0.32em}
    {\small
    \textbf{(b)} Many-particle $U_{\rm NVE}$ block:
    $U_{\rm NVE}(\Delta t)\simeq
    U_1(\Delta t/2)U_2(\Delta t)U_1(\Delta t/2)$.
    }
  \end{minipage}

  \vspace{0.65em}

  \caption{\label{fig:kvn-langevin-circuit}%
    \textbf{Quantum-circuit diagram for the many-particle KvN--Langevin one-step
    evolution.}
    (a)~Overall one-step circuit
    $U_{\mathrm{PITE}}(\sigma_{H})\,U_{\mathrm{F}}(\Delta t)\,
    U_{\mathrm{NVE}}(\Delta t)$ used for canonical state preparation.
    (b)~Implementation of the many-particle $U_{\mathrm{NVE}}(\Delta t)$ block,
    showing the drift--kick--drift structure based on a second-order symmetric
    Suzuki--Trotter decomposition,
    $U_{\mathrm{NVE}}(\Delta t)\simeq
    U_{1}(\Delta t/2)\,U_{2}(\Delta t)\,U_{1}(\Delta t/2)$.
    (c)~PITE/cos-filter implementation of $U_{\mathrm{PITE}}(\sigma_{H})$,
    representing the diffusion block in the momentum Fourier space.}
\end{figure*}

\subsection{\label{sec:quantum-circuit-implementation}Quantum-circuit implementation of the Langevin thermostat}

In the quantum circuit, the $R$ and $P$ registers encode a discretized KvN
amplitude. We denote by $\ket{\psi_n}$ the normalized KvN state after the
$n$th Langevin update. Its amplitudes in the phase-space basis are
$\psi_n(R_j,P_\ell)=\braket{R_j,P_\ell|\psi_n}$, and the corresponding
phase-space density is represented by
\begin{equation}
\begin{aligned}
\rho_n(R_j,P_\ell)
=
|\psi_n(R_j,P_\ell)|^2 .
\end{aligned}
\label{eq:discrete-kvn-density}
\end{equation}
Figure~\ref{fig:kvn-langevin-circuit} shows the circuit structure of one
KvN--Langevin update. Figure~\ref{fig:kvn-langevin-circuit}(a) shows the
one-step sequence composed of $U_{\mathrm{NVE}}$, $U_{\mathrm{F}}$, and
$U_{\mathrm{PITE}}$. Figure~\ref{fig:kvn-langevin-circuit}(b) shows the
drift--kick--drift implementation of $U_{\mathrm{NVE}}$, and
Fig.~\ref{fig:kvn-langevin-circuit}(c) shows the postselected PITE/cos-filter
implementation of the diffusion block.

Combining these components, one KvN--Langevin step is written as
\begin{equation}
\begin{aligned}
\ket{\psi_{n+1}}
=
\frac{
U_{\mathrm{PITE}}(\sigma_H)
U_{\mathrm{F}}(\Delta t)
U_{\mathrm{NVE}}(\Delta t)
\ket{\psi_n}
}{
\left\|
U_{\mathrm{PITE}}(\sigma_H)
U_{\mathrm{F}}(\Delta t)
U_{\mathrm{NVE}}(\Delta t)
\ket{\psi_n}
\right\|
}.
\end{aligned}
\label{eq:one-step-kvn-langevin}
\end{equation}
The normalization factor comes from postselecting the successful branch of the
nonunitary PITE/cos-filter. We now describe the three blocks in the order in
which they appear in Fig.~\ref{fig:kvn-langevin-circuit}(a).

We first consider the NVE propagator. Let $k_R$ and $k_P$ denote the
Fourier-conjugate variables of $R$ and $P$, respectively. Equivalently,
$k_R$ and $k_P$ are the eigenvalues of the differential operators
$\hat{\Lambda}_R=-i\partial_R$ and $\hat{\Lambda}_P=-i\partial_P$ in the
corresponding Fourier bases. We denote the associated diagonal operators by
$\hat{k}_R$ and $\hat{k}_P$. The operator
$\hat{H}_1=\hat{P}\hat{\Lambda}_R/\mu$ is diagonal in the $(k_R,P)$ basis,
whereas $\hat{H}_2=F(\hat{R})\hat{\Lambda}_P$ is diagonal in the $(R,k_P)$
basis. Hence $U_{\mathrm{NVE}}$ can be implemented by quantum Fourier
transforms and diagonal phase gates. We use the second-order symmetric
Suzuki--Trotter decomposition~\cite{Trotter1959,Suzuki1976,suzuki1990}
\begin{equation}
\begin{aligned}
U_{\mathrm{NVE}}(\Delta t)
\simeq
e^{-i(\Delta t/2)\hat{H}_1}
e^{-i\Delta t\hat{H}_2}
e^{-i(\Delta t/2)\hat{H}_1}.
\end{aligned}
\label{eq:nve-suzuki-trotter}
\end{equation}
As illustrated in Fig.~\ref{fig:kvn-langevin-circuit}(b), the corresponding
diagonal phase gates in the one-dimensional notation used here are
\begin{equation}
\begin{aligned}
\Phi_R(\tau)
&=
\exp\!\left[-i\tau\,\frac{\hat{P}\hat{k}_R}{\mu}\right],\\
\Phi_P(\tau)
&=
\exp\!\left[-i\tau\,F(\hat{R})\hat{k}_P\right].
\end{aligned}
\label{eq:phase-gates-nve}
\end{equation}
The gate $\Phi_R$ gives the coordinate drift, and $\Phi_P$ gives the momentum
force kick. The central $\hat{H}_2$ stage, $\Phi_P(\Delta t)$, is therefore the
interface through which the Born--Oppenheimer force from the electronic-structure
calculation enters the nuclear KvN dynamics.

We next implement the friction block. Although friction contracts the momentum
distribution at the density level, it can be represented on the KvN amplitude by
a norm-preserving unitary after symmetrizing the momentum-space dilation. We use
\begin{equation}
\begin{aligned}
\hat{D}_P
&=
\frac{1}{2}
\left(
\hat{P}\hat{\Lambda}_P
+
\hat{\Lambda}_P\hat{P}
\right),
\\
U_{\mathrm F}(\Delta t)
&=
\exp\!\left(+i\gamma\Delta t\,\hat{D}_P\right).
\end{aligned}
\label{eq:U-fric}
\end{equation}
Its action on the KvN amplitude is
\begin{equation}
\begin{aligned}
[U_{\mathrm F}(\Delta t)\psi](R,P)
=
e^{\gamma\Delta t/2}
\psi\!\left(R,e^{\gamma\Delta t}P\right),
\end{aligned}
\label{eq:fric-action}
\end{equation}
which contracts the momentum scale of the corresponding density by
$e^{-\gamma\Delta t}$ per step.

Finally, we describe the diffusion block. The ideal Fokker--Planck diffusion in
Eq.~\eqref{eq:diffusion-substep} acts on the density as a heat kernel in the
momentum direction. In the $k_P$ representation, this ideal density-level
operation multiplies $\widetilde{\rho}(R,k_P,t)$ by
$\exp[-D_P\Delta t\,k_P^2]$, as shown in
Eq.~\eqref{eq:density-heat-kernel}. In the circuit implementation, we
approximate this momentum-space damping by applying a postselected cosine
filter to the KvN amplitude. Specifically, we use the PITE-based operation
\begin{equation}
\begin{aligned}
U_{\mathrm{PITE}}(\sigma_H)
&=
\cos\!\left(\sigma_H\hat{\Lambda}_P\right)
\\
&=
\exp\!\left[
-\frac{\sigma_H^2}{2}\hat{\Lambda}_P^2
-\frac{\sigma_H^4}{12}\hat{\Lambda}_P^4
+ O\!\left((\sigma_H\hat{\Lambda}_P)^6\right)
\right].
\end{aligned}
\label{eq:cos-filter}
\end{equation}
Here $\sigma_H$ is the momentum-scale parameter that controls the strength of
the amplitude-level filter. Equivalently, in the $k_P$ basis, the filter
multiplies each amplitude component by $\cos(\sigma_H k_P)$. Its value is fixed
below by the discrete-time fluctuation--dissipation condition.

As shown in Fig.~\ref{fig:kvn-langevin-circuit}(c), the $P$ register is first
transformed to the $k_P$ basis. Controlled evolutions
$\exp(-i\sigma_H\hat{k}_P)$ and $\exp(+i\sigma_H\hat{k}_P)$ are then applied
with an ancillary qubit, followed by measurement and postselection of the
ancilla onto the success state. This postselected operation realizes the
nonunitary cosine filter $\cos(\sigma_H\hat{\Lambda}_P)$ on the KvN amplitude.
The leading $\hat{\Lambda}_P^2$ term gives Gaussian-like low-pass damping in
momentum Fourier space, whereas the higher-order terms starting from
$\hat{\Lambda}_P^4$ represent deviations from the ideal Gaussian heat kernel.
As discussed below, these higher-order terms lead to a finite time-step
temperature bias.

The friction and diffusion strengths are linked by a discrete-time
fluctuation--dissipation condition. If the diffusion block were an ideal Gaussian
heat kernel retaining only the leading term in Eq.~(\ref{eq:cos-filter}), the
friction step would multiply the mean-square momentum by $e^{-2\gamma\Delta t}$,
whereas the diffusion step would add $\sigma_{H}^{2}/2$ to the density-level
mean-square momentum. Requiring
$\langle P^{2}\rangle=\mu T_{\mathrm{int}}$ to be a fixed point gives
\begin{equation}
  \sigma_{H}^{2}
  =
  2\mu T_{\mathrm{int}}
  \big(1-e^{-2\gamma\Delta t}\big).
  \label{eq:lt-fdt}
\end{equation}
Here $T_{\mathrm{int}}$ is the internal temperature used to set the filter strength.
For an ideal Gaussian diffusion block, Eq.~(\ref{eq:lt-fdt}) is the discrete-time
fluctuation--dissipation relation.

In the actual circuit, however, the Gaussian heat kernel is replaced by the cos
filter. The higher-order terms in
$\log\cos x=-x^{2}/2-x^{4}/12+O(x^{6})$ then accumulate over repeated steps and shift
the equilibrium kinetic temperature. Let
$T_{\mathrm{kin}}\equiv\langle P^{2}\rangle/\mu$ and
$s\equiv\gamma\Delta t$. In the harmonic approximation and continuous-grid limit, the
leading-order bias is
\begin{equation}
  \frac{T_{\mathrm{kin}}-T_{\mathrm{int}}}{T_{\mathrm{int}}}
  =
  \frac{1}{2}\tanh s + O(s^{2}).
  \label{eq:temperature-bias}
\end{equation}
Appendix~\ref{app:cos-bias} gives the derivation. To target the physical temperature
$T_{\mathrm{phys}}$, we therefore choose
\begin{equation}
  T_{\mathrm{int}}
  =
  \frac{T_{\mathrm{phys}}}
  {1+\tfrac{1}{2}\tanh(\gamma\Delta t)}.
  \label{eq:Tint-correction}
\end{equation}
This \textit{a priori} correction cancels the $O(\gamma\Delta t)$ temperature shift
from the cosinefilter in the idealized limit. Residual deviations from finite-grid
effects, anharmonicity, and higher-order terms of the cosine expansion are examined
in Sec.~\ref{sec:canonical} and Appendix~\ref{app:cos-bias}.

By repeatedly applying the one-step circuit in Eq.~(\ref{eq:one-step-kvn-langevin}),
a nonequilibrium KvN state is relaxed toward a canonical nuclear phase-space
distribution.

%==============================================================================
\section{\label{sec:qc-interface}H$_2$ electronic-structure input for the KvN force kick}
%==============================================================================

The construction in Sec.~\ref{sec:framework} was formulated for a general
Born--Oppenheimer potential $V(R)$ and force $F(R)$. We now describe how this
electronic-structure information is supplied to the KvN force-kick block. In
principle, $V(R)$ and $F(R)$ may be obtained from any electronic-structure procedure
that provides a Born--Oppenheimer energy and its nuclear derivative. As a controlled
proof of concept, we use the H$_2$ molecule, for which the relevant electronic
subspace can be reduced to a one-qubit Pauli Hamiltonian.

In the numerical implementation used below, the Hellmann--Feynman force is
precomputed on the $R$ grid and supplied to the KvN circuit as a diagonal phase. This
precompiled implementation allows us to validate the electronic-structure-to-KvN
interface while focusing on canonical state preparation and observable readout.

\begin{figure}[t]
  \centering
  \includegraphics[width=\linewidth]{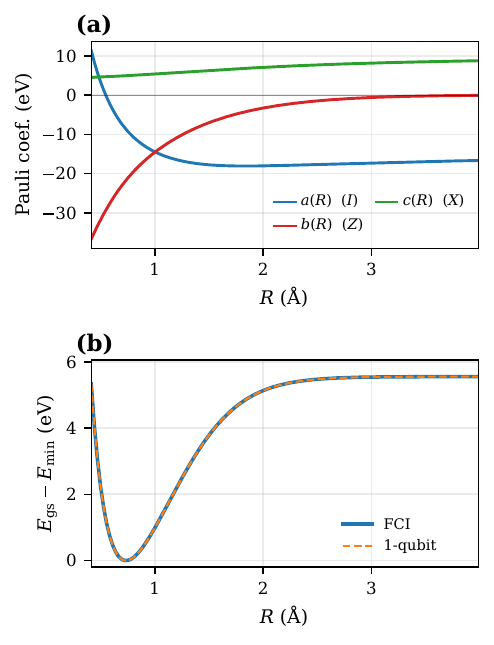}
  \caption{\label{fig:pauli-fci}%
    \textbf{One-qubit Pauli representation of the H$_2$ electronic structure.}
    (a)~Coefficients $a(R)$, $b(R)$, and $c(R)$ in the reduced Hamiltonian
    $\hat{H}_{\mathrm{e}}(R)=a(R)\hat{I}_{e}+b(R)\hat{Z}_{e}+c(R)\hat{X}_{e}$.
    (b)~Ground-state PES relative to its minimum, compared with the FCI/STO-3G
    reference. This representation supplies $V(R)$ and $F(R)$ for the KvN force
    kick used in the molecular-dynamics simulations.}
\end{figure}

Let $\hat{H}_{\mathrm{e}}(R)$ be the electronic Hamiltonian at a fixed internuclear
distance $R$. We use the H$_2$ Hamiltonian in the STO-3G minimal basis and reduce the
electronic subspace containing the ground state to a single qubit through a
fermion-to-qubit transformation and symmetry tapering~\cite{Bravyi2017taper}. Denoting
this electronic qubit by $e$, we write the reduced Hamiltonian as
\begin{equation}
  \hat{H}_{\mathrm{e}}(R)
  =
  a(R)\hat{I}_{e}
  +
  b(R)\hat{Z}_{e}
  +
  c(R)\hat{X}_{e},
  \label{eq:he-pauli}
\end{equation}
where $\hat{I}_{e}$ is the identity operator, and $a(R)$, $b(R)$, and $c(R)$ are real
coefficients determined by the electronic integrals. The internuclear distance $R$
acts as a parameter that changes the electronic Hamiltonian. The $R$ dependence of
these coefficients is shown in Fig.~\ref{fig:pauli-fci}(a).

For each fixed $R$, Eq.~(\ref{eq:he-pauli}) is a $2\times2$ matrix, so its
ground-state energy is obtained analytically as
\begin{equation}
  E_{\mathrm{gs}}(R)
  =
  a(R)-\Omega(R),
  \qquad
  \Omega(R)=\sqrt{b(R)^{2}+c(R)^{2}}.
  \label{eq:Egs-pauli}
\end{equation}
We use this ground-state energy as the Born--Oppenheimer potential,
\begin{equation}
  V(R)=E_{\mathrm{gs}}(R).
  \label{eq:V-from-Egs}
\end{equation}
An overall energy shift does not affect the nuclear dynamics. Therefore,
Fig.~\ref{fig:pauli-fci}(b) shows the potential-energy surface (PES) relative to its
minimum, $E_{\mathrm{gs}}(R)-E_{\mathrm{min}}$. The same panel confirms that the
one-qubit expression in Eq.~(\ref{eq:Egs-pauli}) reproduces the direct FCI/STO-3G
reference.

The force used in KvN molecular dynamics is the negative derivative of this
Born--Oppenheimer potential. By the Hellmann--Feynman theorem,
\begin{equation}
  D_{R}^{\mathrm{el}}(R)
  =
  \bra{g(R)}\partial_{R}\hat{H}_{\mathrm{e}}(R)\ket{g(R)}
  =
  \partial_{R}E_{\mathrm{gs}}(R),
  \label{eq:Del-pauli}
\end{equation}
where $\ket{g(R)}$ is the ground state of Eq.~(\ref{eq:he-pauli}). For the one-qubit
Hamiltonian in Eq.~(\ref{eq:he-pauli}), this derivative becomes
\begin{equation}
  D_{R}^{\mathrm{el}}(R)
  =
  \partial_{R}a(R)
  -
  \frac{
  b(R)\partial_{R}b(R)
  +
  c(R)\partial_{R}c(R)
  }{\Omega(R)}.
  \label{eq:Del-pauli-explicit}
\end{equation}
Thus the physical force is
\begin{equation}
  \begin{aligned}
  F(R)
  &=
  -D_{R}^{\mathrm{el}}(R)\\
  &=
  -\partial_{R}a(R)
  +
  \frac{
  b(R)\partial_{R}b(R)
  +
  c(R)\partial_{R}c(R)
  }{\Omega(R)}.
  \end{aligned}
  \label{eq:hf-force-pauli}
\end{equation}

This force enters the KvN propagator through the force-kick term
$\hat{H}_{2}=F(\hat{R})\hat{\Lambda}_{P}$. In the Fourier-transformed momentum basis,
the corresponding stage of the NVE update is the $R$-controlled diagonal phase
\begin{equation}
  \exp\!\big[-i\Delta t\,F(\hat{R})\,\hat{k}_{P}\big],
  \label{eq:force-phase}
\end{equation}
where $\hat{k}_{P}$ is the diagonal operator representing the eigenvalues of
$\hat{\Lambda}_{P}$. On the discrete $R$ grid used in the state-vector simulations,
this phase acts as
\begin{equation}
  \ket{R_{j},k_{P,\ell}}
  \longmapsto
  \exp[-i\Delta t\,F(R_{j})\,k_{P,\ell}]
  \ket{R_{j},k_{P,\ell}}.
  \label{eq:compiled-force-kick}
\end{equation}
Thus, the precomputed Hellmann--Feynman force is embedded directly as the KvN
force-kick phase.

This implementation should be distinguished from an on-the-fly electronic-register
construction. Huang \textit{et al.} proposed using the Hellmann--Feynman derivative
as a nuclear-coordinate-controlled diagonal block encoding and combining it with a
momentum-derivative operator to construct an electronic Liouvillian
~\cite{Huang2025alchemist}. Equation~(\ref{eq:compiled-force-kick}) is a precompiled
one-dimensional counterpart of the same force-kick idea for the one-qubit H$_2$
model. We do not keep an explicit electronic register during the KvN evolution.
Instead, we precompute $F(R)=-D_{R}^{\mathrm{el}}(R)$ on the $R$ grid and use the
resulting diagonal phase to validate the electronic-structure-to-KvN interface. This
choice removes the electronic register from the KvN evolution, reduces the active
qubit count, and allows us to focus on canonical state preparation and observable
readout.

The following numerical demonstrations use this H$_2$ PES and the precomputed
Hellmann--Feynman force kick to test canonical state preparation, the Raman-active
VDOS readout, and the TST flux expectation.

%==============================================================================
\section{\label{sec:canonical}Canonical state preparation and bond-formation dynamics}
%==============================================================================
\begin{figure*}[t]
  \centering
  \includegraphics[width=0.95\textwidth]{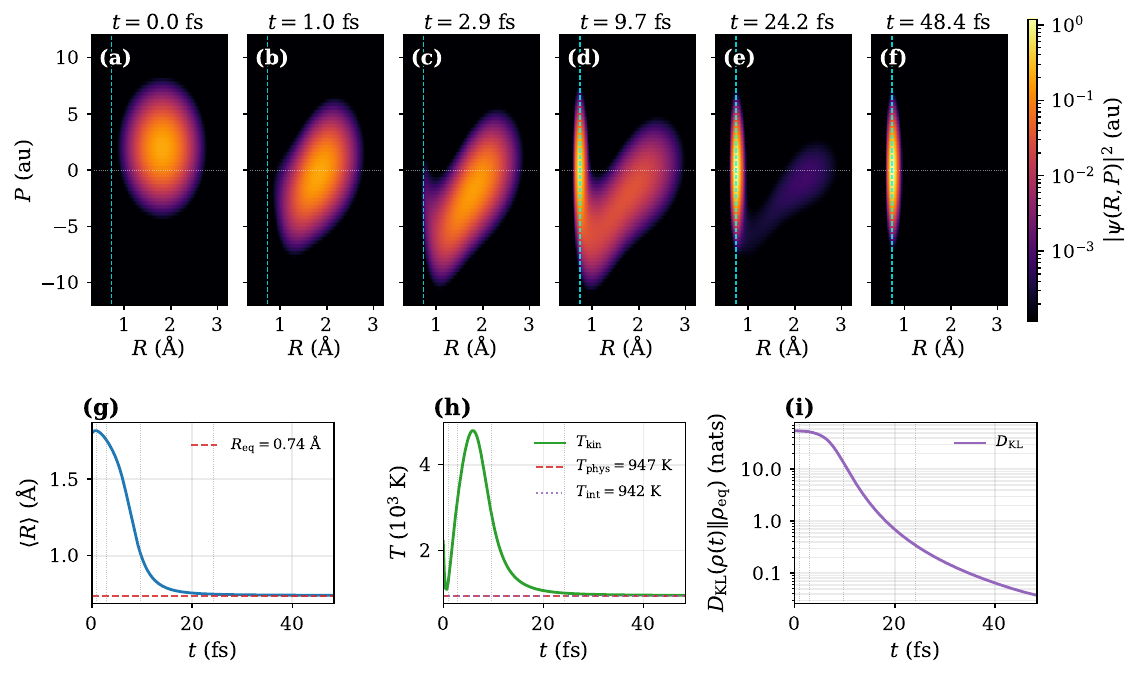}
  \caption{\label{fig:canonical}%
    \textbf{Canonical state preparation and relaxation toward the H$_2$ bond.}
    (a)--(f)~Log-scale snapshots of the phase-space density
    $|\psi(R,P,t)|^{2}$. An initial wave packet on the dissociation side moves
    toward the bond-length region and relaxes under the KvN--Langevin thermostat.
    (g)~Mean internuclear distance $\langle R\rangle(t)$.
    (h)~Kinetic temperature $T_{\mathrm{kin}}(t)$, with
    $T_{\mathrm{phys}}=947$~K and the corrected internal temperature
    $T_{\mathrm{int}}=942$~K.
    (i)~Kullback--Leibler divergence
    $D_{\mathrm{KL}}(\rho(t)\|\rho_{\mathrm{eq}})$ from the canonical reference
    distribution.}
\end{figure*}

We now apply the KvN--Langevin one-step circuit of Sec.~\ref{sec:framework} to the
H$_2$ Born--Oppenheimer potential and validate canonical phase-space state
preparation. This validation has two aims. First, we test whether the Langevin
thermostat, including the PITE/cos-filter diffusion block, relaxes a strongly
nonequilibrium initial state toward a finite-temperature canonical distribution.
Second, by following an initial nuclear wave packet placed on the dissociation side,
we check that the electronic-structure-derived PES and force introduced in
Sec.~\ref{sec:qc-interface} drive the KvN nuclear dynamics consistently toward the
bonding region. All results in this section are obtained by applying
Eq.~(\ref{eq:one-step-kvn-langevin}) as a state-vector operation on a discrete grid;
device noise and finite-shot statistics are not included.

The nuclear coordinate and momentum are represented by registers with
$n_{R}=n_{P}=7$ qubits, corresponding to uniform grids with
$N_{R}=N_{P}=128$ points. The coordinate range is chosen to include both the H$_2$
bonding region and the dissociation-side initial state. The potential and force are
computed from the FCI/STO-3G PES represented by the one-qubit Pauli encoding
described in Sec.~\ref{sec:qc-interface}. Unless otherwise stated, we use
$T_{\mathrm{phys}}=947$~K, $\gamma=0.02$~a.u., and
$\Delta t=0.5$~a.u. These parameters give
$s=\gamma\Delta t=0.01$, and Eq.~(\ref{eq:Tint-correction}) gives the corrected
internal temperature $T_{\mathrm{int}}\simeq 942$~K. The cos-filter strength
$\sigma_{H}$ is then determined from Eq.~(\ref{eq:lt-fdt}).

As the initial state, we use a normalized Gaussian wave packet localized on the
dissociation side, with center $R_{0}=1.82$~\AA{} and $P_{0}=0$, and with finite
widths in both the $R$ and $P$ directions. This setup allows us to monitor how the
internuclear distribution moves down the PES toward the bond-length region and then
thermalizes under the Langevin thermostat.

During the time evolution, the phase-space density is
$\rho_{t}(R,P)=|\psi_{t}(R,P)|^{2}$. We monitor relaxation using three diagnostics.
The first is the mean internuclear distance,
\begin{equation}
  \langle R\rangle_{t} = \int dR\,dP\; R\,\rho_{t}(R,P),
  \label{eq:R-mean-canonical}
\end{equation}
which tracks motion from the dissociation side to the bonding region. The second is
the kinetic temperature,
\begin{equation}
  T_{\mathrm{kin}}(t)
  = \frac{\langle P^{2}\rangle_{t}}{\mu}
  = \frac{1}{\mu}\int dR\,dP\; P^{2}\rho_{t}(R,P),
  \label{eq:Tkin-canonical}
\end{equation}
which probes thermalization of the momentum distribution. The third is the
Kullback--Leibler divergence from the canonical reference distribution,
\begin{equation}
  \rho_{\mathrm{eq}}(R,P)
  = \frac{1}{Z}\exp\!\left[-\frac{1}{T_{\mathrm{phys}}}
    \left(\frac{P^{2}}{2\mu} + V(R)\right)\right],
  \label{eq:rhoeq-canonical}
\end{equation}
defined as
\begin{equation}
  D_{\mathrm{KL}}(\rho_{t}\|\rho_{\mathrm{eq}})
  = \int dR\,dP\;\rho_{t}(R,P)
    \log\frac{\rho_{t}(R,P)}{\rho_{\mathrm{eq}}(R,P)}.
  \label{eq:DKL-canonical}
\end{equation}
Here $Z$ is the normalization constant. The first two quantities characterize the
relaxation of representative position and momentum moments, whereas
$D_{\mathrm{KL}}$ measures the approach of the full phase-space distribution to the
canonical reference.

Figure~\ref{fig:canonical}(a)--(f) shows logarithmic-scale snapshots of the
phase-space density from $t=0$ to $48.4$~fs. Initially, the density is localized on
the dissociation side. By $t\simeq 1.0$~fs, the PES gradient has generated
negative-momentum components, producing a tilted distribution with correlations
between $R$ and $P$. At $t\simeq 2.9$~fs, the density has flowed further into the
bonding region. At later times, a residual nonequilibrium tail remains on the
dissociation side, while the dominant component becomes localized near the
equilibrium bond length. By the final time, the main density is concentrated near
$R_{\mathrm{eq}}$ and is also close to thermalized in the momentum direction. These
snapshots show relaxation from a dissociation-side wave packet toward a canonical
phase-space distribution around the H$_2$ bond.

Figure~\ref{fig:canonical}(g) shows that $\langle R\rangle_{t}$ relaxes from the
initial value $1.82$~\AA{} to the equilibrium bond length
$R_{\mathrm{eq}}\simeq 0.74$~\AA{}. Figure~\ref{fig:canonical}(h) shows the kinetic
temperature. The initial potential energy is first converted into kinetic energy,
and $T_{\mathrm{kin}}(t)$ reaches approximately $4.9\times 10^{3}$~K near
$t\simeq 5$~fs. It then decreases through the combined action of friction and
diffusion and relaxes to a temperature close to the target value. The horizontal
lines indicate $T_{\mathrm{phys}}=947$~K and the corrected internal temperature
$T_{\mathrm{int}}\simeq 942$~K; their difference comes from the leading-order
cos-filter correction in Eq.~(\ref{eq:Tint-correction}).

Figure~\ref{fig:canonical}(i) shows the time evolution of
$D_{\mathrm{KL}}(\rho_{t}\|\rho_{\mathrm{eq}})$. The initial value is about
$30$, reflecting the strong mismatch between the localized initial wave packet
and the canonical distribution. During the evolution, $D_{\mathrm{KL}}$ decreases
almost monotonically and reaches approximately $0.05$~nats at $48.4$~fs. The
remaining deviation is expected because of finite evolution time, grid
discretization, the cos-filter approximation, and anharmonicity of the PES.
Nevertheless, the reduction by nearly three orders of magnitude quantitatively
demonstrates relaxation toward the canonical reference distribution.

These results show that the KvN--Langevin/PITE circuit prepares a canonical
phase-space state under the controlled H$_2$ conditions considered here. They also
verify that the PES and Hellmann--Feynman force obtained from the electronic-structure
calculation can be passed consistently to the KvN nuclear dynamics. In the next
section, we use the canonical KvN state prepared here for a dynamical readout of the
VDOS associated with the Raman-active H--H stretch coordinate.

%==============================================================================
\section{\label{sec:vdos}Vibrational density of states of a Raman-active mode}
%==============================================================================

Starting from the canonical KvN state prepared in Sec.~\ref{sec:canonical}, we now
read out the vibrational density of states (VDOS) associated with the H--H stretch
coordinate. The state preparation uses the KvN--Langevin procedure described above,
whereas the spectral readout uses the thermostat-free KvN time evolution generated
by $\hat{H}_{\mathrm{NVE}}$ in Eq.~\eqref{eq:H-nve}.

We first define the KvN observable. The centered H--H stretch coordinate is
\begin{equation}
\begin{aligned}
\hat{Q}
\equiv
\hat{R}
-
\big\langle \hat{R} \big\rangle_{\mathrm{eq}},
\end{aligned}
\label{eq:Q-def}
\end{equation}
where $\big\langle \hat{R} \big\rangle_{\mathrm{eq}}$ is the mean internuclear
distance in the canonical KvN state. The VDOS is obtained from the equilibrium
autocorrelation function
\begin{equation}
\begin{aligned}
C_{QQ}(t)
=
\bra{\psi_{\mathrm{eq}}}
\hat{Q}\,
e^{-it\hat{H}_{\mathrm{NVE}}}\,
\hat{Q}
\ket{\psi_{\mathrm{eq}}},
\end{aligned}
\label{eq:vdos-correlation}
\end{equation}
and its Fourier spectrum,
\begin{equation}
\begin{aligned}
S_{QQ}(\omega)
=
\int_{-\infty}^{\infty} dt\; e^{i\omega t} C_{QQ}(t).
\end{aligned}
\label{eq:vdos-spectrum}
\end{equation}
The thermostat-free KvN Hamiltonian $\hat{H}_{\mathrm{NVE}}$ is the
amplitude-level generator defined in Eq.~\eqref{eq:H-nve}. The canonical KvN
state $\ket{\psi_{\mathrm{eq}}}$ satisfies
$|\psi_{\mathrm{eq}}(R,P)|^{2}=\rho_{\mathrm{eq}}(R,P)$.

To resolve the positive- and negative-frequency components in QPE, we prepare
branch-selective excitation states before applying the QPE circuit. In this
work, branch-selective QPE denotes a standard QPE readout whose input state is
chosen to enhance either the positive- or negative-frequency component of the
KvN spectrum. For this purpose, we use
\begin{equation}
\begin{aligned}
\hat{A}_{\pm}
=
\hat{Q}
\mp
i\hat{\Pi},
\qquad
\hat{\Pi}
\equiv
\frac{\hat{P}}{\mu\omega_{\mathrm{ref}}},
\end{aligned}
\label{eq:Apm-def}
\end{equation}
rather than $\hat{Q}$ itself. The parameter $\omega_{\mathrm{ref}}$ is a
reference frequency used to make $\hat{Q}$ and
$\hat{P}/(\mu\omega_{\mathrm{ref}})$ have comparable units and to improve the
separation of the positive- and negative-frequency branches. In the present
H$_2$ demonstration, we choose $\omega_{\mathrm{ref}}$ from the harmonic
curvature of the Born--Oppenheimer PES near the equilibrium bond length,
\begin{equation}
\begin{aligned}
\omega_{\mathrm{ref}}
=
\sqrt{
\frac{1}{\mu}
\left.
\frac{d^{2}V(R)}{dR^{2}}
\right|_{R=R_{\mathrm{eq}}}
}.
\end{aligned}
\label{eq:omega-ref-def}
\end{equation}
In the numerical implementation, this curvature is obtained from a local
quadratic fit to the FCI/STO-3G PES used in the KvN dynamics.

The reference frequency $\omega_{\mathrm{ref}}$ does not determine the QPE
frequency bins or fix the spectral peak position. The frequencies read out by
QPE are obtained from the eigenphases of the thermostat-free KvN evolution
generated by $\hat{H}_{\mathrm{NVE}}$. An inaccurate choice of
$\omega_{\mathrm{ref}}$ therefore reduces the separation between the
positive- and negative-frequency branches, but it does not shift the
eigenfrequencies obtained from the QPE readout.

In the harmonic approximation, $\hat{A}_{+}$ and $\hat{A}_{-}$ predominantly
select the positive- and negative-frequency branches, respectively, when
$\omega_{\mathrm{ref}}$ is close to the local vibrational frequency. These
branches are analogous to the frequency components that would be associated with
Stokes and anti-Stokes processes in a Raman setting. However, the branch weights
obtained in this readout are not Raman intensities. A Raman intensity
calculation would require an electronic response factor, in particular the
nuclear-coordinate-dependent electronic polarizability. In classical KvN
theory, $\hat{Q}$ and $\hat{\Pi}$ commute. Therefore,
\begin{equation}
\begin{aligned}
\hat{A}_{+}^{\dagger}\hat{A}_{+}
=
\hat{A}_{-}^{\dagger}\hat{A}_{-}
=
\hat{Q}^{2}
+
\hat{\Pi}^{2}.
\end{aligned}
\label{eq:Apm-equal-norm}
\end{equation}
The integrated branch weights are therefore symmetric in the present KvN
amplitude readout.

Figure~\ref{fig:lcu-qpe} shows the readout circuit. In
Fig.~\ref{fig:lcu-qpe}(a), a linear-combination-of-unitaries (LCU) operation
$U_{\alpha}$ prepares a flagged superposition of the normalized
branch-selective states
$\ket{\alpha_{\pm}}
=
\hat{A}_{\pm}\ket{\psi_{\mathrm{eq}}}/
\|\hat{A}_{\pm}\ket{\psi_{\mathrm{eq}}}\|$.
The prepared state is then used as the input to QPE, as shown in
Fig.~\ref{fig:lcu-qpe}(b). The QPE circuit applies controlled powers of
\begin{equation}
\begin{aligned}
U_{\mathrm{NVE}}(\tau)
=
\exp\!\left[-i\tau \hat{H}_{\mathrm{NVE}}\right],
\end{aligned}
\end{equation}
where $\hat{H}_{\mathrm{NVE}}$ is defined in Eq.~\eqref{eq:H-nve}. Measuring the
phase ancilla qubits together with the flag gives the QPE distributions
conditioned on the positive- and negative-frequency branches. The results below
are obtained from state-vector simulations of the corresponding circuit.

\begin{figure}[t]
  \centering
  \includegraphics[width=\linewidth]{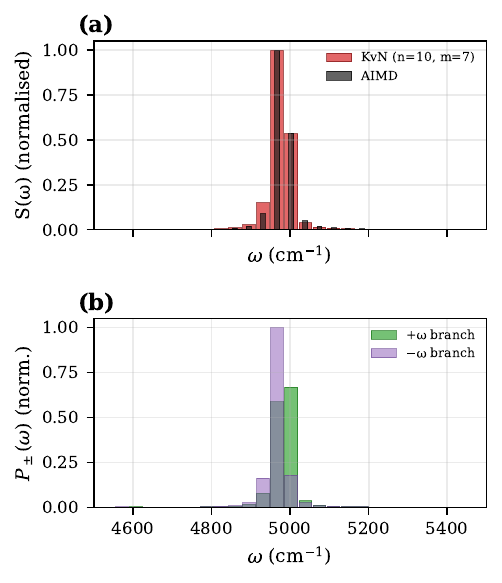}
  \caption{\label{fig:vdos}%
    \textbf{VDOS readout for the Raman-active H--H stretch coordinate.}
    (a)~KvN--QPE spectrum compared with the AIMD reference on the same FCI/STO-3G
    PES, with the AIMD spectrum matched to the QPE time window and binning.
    (b)~Branch-resolved QPE output obtained from $\hat{A}_{+}$ and $\hat{A}_{-}$ and
    overlaid on the $|\omega|$ axis. The branch weights are symmetric in classical
    KvN theory and should not be interpreted as Raman Stokes or anti-Stokes
    intensities, because the electronic polarizability is not included.}
\end{figure}

Figure~\ref{fig:vdos} shows the VDOS readout for H$_2$. We use grids with
$n_R=n_P=10$, set the number of QPE ancilla qubits to $m=7$, and choose
$T_{\mathrm{phys}}=300$~K. For comparison, we use the AIMD reference defined in
Appendix~\ref{app:aimd-postprocess}, postprocessed to match the finite time
window and frequency binning of QPE.

Figure~\ref{fig:vdos}(a) compares the KvN--QPE spectrum with the AIMD
reference. The two spectra peak in the same QPE frequency bin. Under the
plotted conditions, the corresponding bin center is $4966.85$~cm$^{-1}$ for
both spectra, with no resolved difference at the displayed bin resolution. This
agreement confirms, within the present benchmark, that the QPE readout recovers
the VDOS peak obtained from the corresponding trajectory-based reference.

Figure~\ref{fig:vdos}(b) shows the raw branch-resolved QPE output obtained from
$\hat{A}_{+}$ and $\hat{A}_{-}$. The flag distinguishes the two branches, which
are overlaid on the $|\omega|$ axis for comparison. The peaks appear near the
same vibrational frequency, indicating that the LCU-based branch-selective
preparation separates the two frequency components as intended. The nearly
symmetric integrated branch weights are consistent with
Eq.~\eqref{eq:Apm-equal-norm}) and reflect the classical KvN amplitude
construction, not a Raman Stokes/anti-Stokes intensity ratio.

These results show that the canonical KvN state prepared in
Sec.~\ref{sec:canonical} can be used not only as a static equilibrium
distribution but also as an input state for dynamical readout based on time
correlations.

\begin{figure}[t]
  \centering
  \small

  {\small\textbf{(a)} LCU state preparation for branch-selective excitation.}
  \vspace{0.35em}
  \begin{adjustbox}{max width=\columnwidth,center}
  \begin{quantikz}[column sep=14pt, row sep=18pt]
    \lstick{$|\psi_{\mathrm{eq}}\rangle$ \\ \scriptsize($n_R+n_P$ q)}
      & \qwbundle{}
      & \gate[wires=2]{\;U_{\alpha}\;}
      & \qw
      \rstick{$\dfrac{|\alpha_+\rangle|0\rangle_f + |\alpha_-\rangle|1\rangle_f}{\sqrt{2}\,\lambda}$} \\
    \lstick{$|0\rangle_f$ \\ \scriptsize(flag, 1 q)}
      & \qw
      & \ghost{\;U_{\alpha}\;}
      & \qw
  \end{quantikz}
  \end{adjustbox}

  \vspace{0.9em}

  {\small\textbf{(b)} Quantum phase estimation with a frequency shift.}
  \vspace{0.35em}
  \begin{adjustbox}{max width=\columnwidth,center}
  \begin{quantikz}[column sep=10pt, row sep=14pt]
    \lstick{$|\alpha_\pm\rangle$\ \scriptsize($n_R{+}n_P$ q)}
      & \qwbundle{}
      & \gate{U^{2^0}}
      & \gate{U^{2^1}}
      & \cdots
      & \gate{U^{2^{m-1}}}
      & \qw
      & \qw
      & \qw \\
    \lstick{$a_0:\ |0\rangle$}
      & \qw
      & \gate{H}
      & \ctrl{-1}
      & \qw
      & \qw
      & \qw
      & \gate[wires=3]{e^{+\mathrm{i}\omega_{\mathrm{shift}} k\tau}}
      & \gate[wires=3]{\mathrm{IQFT}_m}
      & \meter{} \\
    \lstick{$a_1:\ |0\rangle$}
      & \qw
      & \gate{H}
      & \qw
      & \ctrl{-2}
      & \qw
      & \qw
      & \ghost{e^{+\mathrm{i}\omega_{\mathrm{shift}} k\tau}}
      & \ghost{\mathrm{IQFT}_m}
      & \meter{} \\
    \lstick{$a_{m-1}:\ |0\rangle$}
      & \qw
      & \gate{H}
      & \qw
      & \qw
      & \ctrl{-3}
      & \qw
      & \ghost{e^{+\mathrm{i}\omega_{\mathrm{shift}} k\tau}}
      & \ghost{\mathrm{IQFT}_m}
      & \meter{}
  \end{quantikz}
  \end{adjustbox}

  \caption{\label{fig:lcu-qpe}%
    \textbf{LCU+QPE circuit for the VDOS readout.}
    (a)~The LCU unitary $U_{\alpha}$ prepares the branch-selective states
    $\ket{\alpha_{\pm}}\propto\hat{A}_{\pm}\ket{\psi_{\mathrm{eq}}}$ with a flag
    qubit.
    (b)~QPE is applied to $\ket{\alpha_{\pm}}$ using the Hamiltonian KvN time
    evolution, and the flag and phase ancilla measurements give the
    branch-resolved frequency distributions.}
\end{figure}

%==============================================================================
\section{\label{sec:tst}Transition-state-theory rate as a static canonical observable}
%==============================================================================

We next evaluate the transition-state-theory (TST) rate constant along the H$_2$
dissociation coordinate as a static observable of the canonical distribution. In
Sec.~\ref{sec:vdos}, the VDOS was read out as a dynamical
spectrum using Hamiltonian KvN time evolution and QPE. In
contrast, TST expresses the rate constant as the ratio of an equilibrium
dividing-surface flux to a reactant population. The readout in this section therefore
does not require a time-evolution circuit. It only requires expectation values of
diagonal observables over the canonical phase-space distribution.

We take the H--H internuclear distance $R$ as the reaction coordinate and place the
dividing surface at $R=R^{\ddagger}$. The TST rate constant is written as
\begin{equation}
  k_{\mathrm{TST}}(T)
  = \frac{\Phi^{\ddagger}_{+}(T)}{P_{\mathrm{R}}(T)},
  \label{eq:tst-rate}
\end{equation}
where $T$ is the temperature, $\Phi^{\ddagger}_{+}$ is the canonical average of the
positive flux across the dividing surface in the dissociation direction, and
$P_{\mathrm{R}}$ is the canonical probability of the reactant region. We evaluate
\begin{align}
  \Phi^{\ddagger}_{+}(T)
  &=
  \int dR\,dP\;
  \delta_{\sigma}(R-R^{\ddagger})\,
  \Theta(P)\,
  \frac{P}{\mu}\,
  \rho_{\mathrm{eq}}(R,P;T),
  \label{eq:tst-flux}\\
  P_{\mathrm{R}}(T)
  &=
  \int dR\,dP\;
  \Theta(R^{\ddagger}-R)\,
  \rho_{\mathrm{eq}}(R,P;T),
  \label{eq:tst-reactant}
\end{align}
where $\delta_{\sigma}$ is a smoothed delta function of width $\sigma$ representing
the dividing surface on the grid, $\Theta$ is the Heaviside step function. Both quantities are diagonal in the computational basis of
$R$ and $P$. Thus, once a KvN state satisfying
$|\psi_{\mathrm{eq}}(R,P;T)|^{2}=\rho_{\mathrm{eq}}(R,P;T)$ is available, the flux
and population can be evaluated as grid sums or, on a quantum device, as diagonal
expectation values.

We evaluate the TST estimator on a canonical KvN state, which is the equilibrium
target of the KvN--Langevin preparation in Sec.~\ref{sec:framework}. In the numerical
benchmark, this state is analytically encoded on the H$_2$ grid as
\begin{equation}
  \psi_{\mathrm{eq}}(R_{j},P_{\ell};T)
  \propto
  \exp\!\left[
  -\frac{1}{2T}
  \left(
  \frac{P_{\ell}^{2}}{2\mu}+V(R_{j})
  \right)
  \right],
  \label{eq:analytic-canonical-state}
\end{equation}
and Eqs.~(\ref{eq:tst-flux}) and (\ref{eq:tst-reactant}) are evaluated from
$|\psi_{\mathrm{eq}}|^{2}$. Details of the grid estimator are given in
Appendix~\ref{app:tst-details}.

Figure~\ref{fig:tst-rate} shows $k_{\mathrm{TST}}$ for
$T=2500$, $5000$, and $10000$~K in an Arrhenius representation. The green points are
obtained by applying the flux estimator to the analytically encoded canonical KvN
state vector. The gray line is a guide to the eye and is used to indicate the
Arrhenius-type temperature dependence. The smooth behavior of the green points shows
that the canonical-state flux estimator remains well defined over the temperature
range considered here.

For comparison, Fig.~\ref{fig:tst-rate} also shows a particle-based crossing
reference. We first sample canonical initial conditions and then propagate
finite-time NVE trajectories of length $t_{\mathrm{sim}}$. The reference rate is
estimated by counting positive crossings of the dividing surface. At high
temperature, such crossings are sampled. At low temperature, however, no crossing is
observed for the same number and length of trajectories, so the estimate is limited
by the zero-count detection floor indicated by the horizontal dashed line. In
contrast, the KvN state-vector calculation evaluates $k_{\mathrm{TST}}$ directly as a
static flux expectation over the canonical distribution. This trajectory-counting
floor is therefore absent at the level of the state-vector flux evaluation.

\begin{figure}[t]
  \centering
  \includegraphics[width=\linewidth]{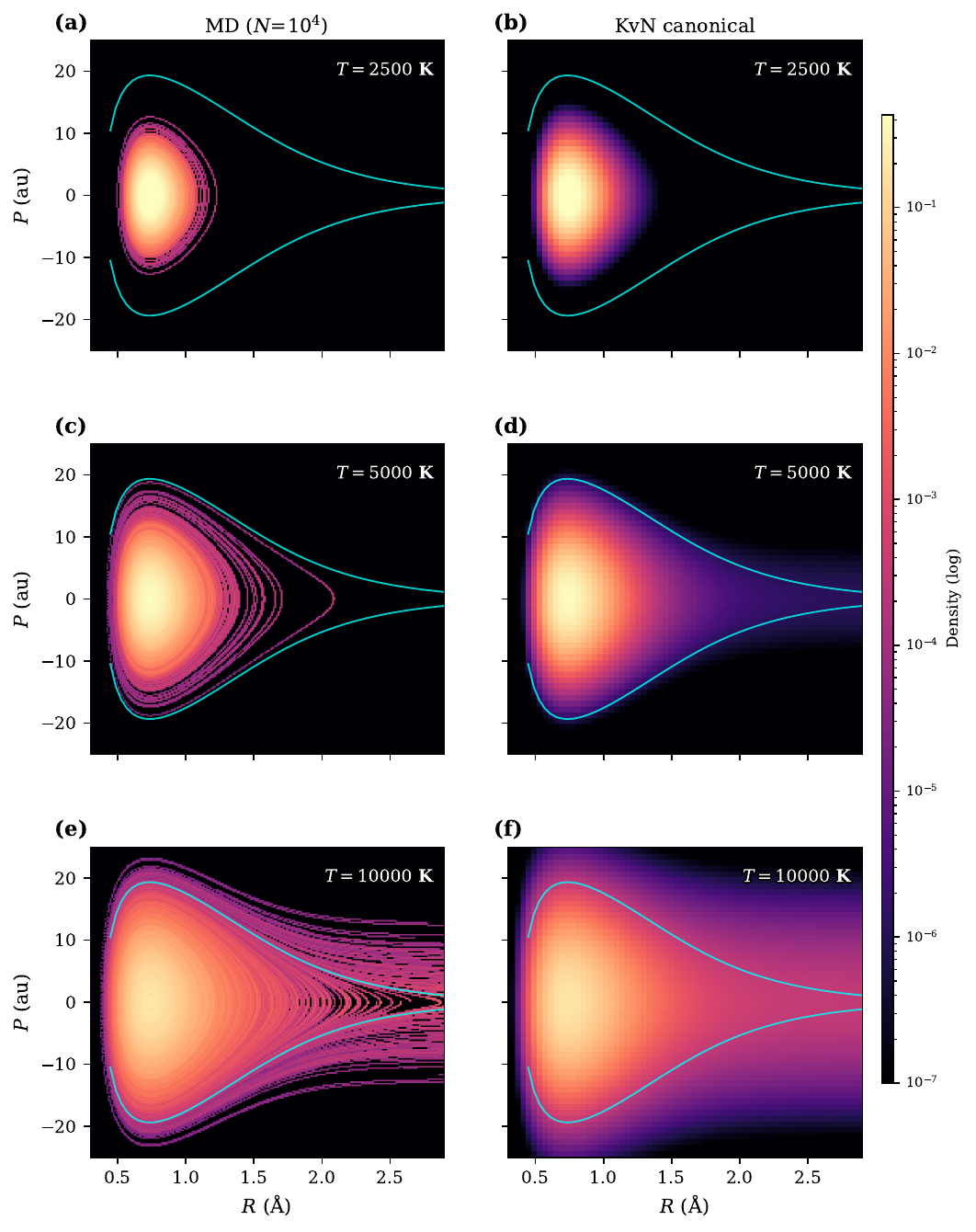}
  \caption{\label{fig:phase-space}%
    \textbf{Phase-space densities in the TST rare-event region.}
    The left column shows density estimates from finite-particle samples drawn from
    the canonical distribution. The right column shows the analytically encoded
    canonical KvN density on the grid. At low temperature, finite-particle samples
    become sparse near the dividing surface, whereas the canonical KvN encoding
    retains the Boltzmann-suppressed density over the full grid.}
\end{figure}

Figure~\ref{fig:phase-space} visualizes the same contrast in phase space. The left
column shows density estimates from a finite number of particle samples drawn from
the canonical distribution. The right column shows the canonical
KvN density on the grid. At high temperature, the finite-particle samples represent
the region near the dividing surface to some extent. At low temperature, the sampled
density in the reactive region becomes sparse. The canonical KvN representation, in
contrast, retains the Boltzmann-suppressed density near the dividing surface smoothly
over the full grid. This comparison illustrates how representing the full canonical
distribution as a state differs from estimating rare regions by finite-particle
sampling.

The state-vector results in this section should not be interpreted as hardware
sampling costs. If the static flux were measured by simple computational-basis
sampling on a quantum device, estimating the contribution from small-probability
regions would still require many shots. Nevertheless, Eq.~(\ref{eq:tst-flux}) is a
nonnegative weighted expectation value in the computational basis. If coherent
canonical state preparation and the corresponding weight oracle are available, this
observable can in principle be combined with amplitude-estimation-type
expectation-value estimation.

These results show that, within the TST framework, the reaction rate can be evaluated
in the KvN representation as a static observable of a canonical state. This contrasts
with the VDOS readout in Sec.~\ref{sec:vdos}, where the observable is dynamical and
is extracted from the eigenfrequencies of Hamiltonian KvN time evolution. The two
examples demonstrate that the same canonical KvN state can serve as a common input
for either dynamical or static readout protocols, depending on the observable.

\begin{figure}[t]
  \centering
  \includegraphics[width=\linewidth]{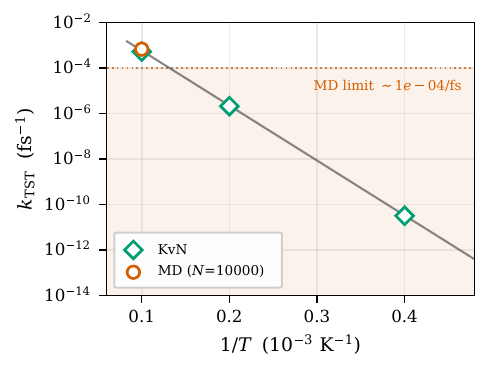}
  \caption{\label{fig:tst-rate}%
    \textbf{Arrhenius representation of the TST rate for H$_2$ dissociation.}
    The green points are obtained by applying the flux estimator to the analytically
    encoded canonical KvN state vector. The gray line is a guide to the eye for the
    Arrhenius-type trend and is not an independent theory curve. The orange points
    show a particle-based crossing reference from finite-time NVE trajectories
    initialized from the canonical distribution. The horizontal dashed line indicates
    the effective detection limit under the same sampling conditions.}
\end{figure}

%==============================================================================
\section{\label{sec:conclusion-outlook}Conclusion and outlook}
%==============================================================================

In summary, we constructed a quantum-circuit framework for preparing a canonical
nuclear phase-space state by representing the classical nuclear distribution as a
KvN wave function and incorporating a Langevin thermostat under the
Born--Oppenheimer approximation. The central component is a KvN--Langevin one-step
update that combines Hamiltonian Liouville flow, momentum friction, and a
PITE/cos-filter momentum-diffusion block. We derived an analytic leading-order
correction for the temperature bias introduced by the higher-order terms of the
cosinefilter, and verified in state-vector simulations for H$_2$ that a strongly
nonequilibrium phase-space distribution relaxes toward the canonical distribution.

The H$_2$ electronic-structure input provides a minimal molecular test of the
electronic-structure-to-KvN interface. In the numerical implementation, the
Born--Oppenheimer potential and the Hellmann--Feynman force obtained from the
one-qubit Pauli Hamiltonian were precomputed on the $R$ grid, and the force was
embedded as the diagonal phase of the KvN force kick. This precompiled force kick
reduces the active qubit count and allows the present demonstrations to focus on
canonical state preparation and observable readout, while retaining the same
force-kick structure that an $R$-controlled electronic-force routine would generate.

Starting from the canonical KvN state, we demonstrated two complementary readouts.
First, we evaluated the VDOS associated with the Raman-active H--H stretch coordinate
using branch-selective state preparation and QPE. The KvN--QPE peak agrees with the
AIMD reference within the same finite time window and frequency binning, showing that
the Hamiltonian KvN evolution captures the vibrational frequency of the classical
Liouville dynamics. Second, within transition-state theory, we evaluated the
dissociation rate as a static flux expectation over the canonical distribution. This
shows that the TST rate can be treated as a diagonal observable of the equilibrium
phase-space distribution, rather than being estimated from rare crossings in
finite-time trajectories.

A key next step is the extension to systems with many nuclear degrees of freedom.
For H$_2$, the Born--Oppenheimer potential and force can be precomputed on a
one-dimensional $R$ grid, whereas a full force table does not scale for realistic
molecules and materials. Scalable implementations will therefore require efficient
potential and force oracles, compressed representations, or on-the-fly force
construction. Importantly, the KvN force-kick block only requires access to
$V(\mathbf{R})$ and $\nabla V(\mathbf{R})$ and is not restricted to electronic
structure calculations. The same interface could also be connected to rapidly
developing machine-learning potentials, which provide compact surrogate models for
Born--Oppenheimer energies and forces.

The present framework therefore opens a possible route to evaluating
rare-event-driven phenomena in realistic materials directly from canonical
phase-space states, with future progress in canonical state preparation, force
construction, and measurement primitives providing the key ingredients for scaling
this approach. Examples include activated diffusion, defect migration, and reaction
events in complex materials. Taken together, this work establishes a practical end-to-end quantum-circuit pipeline
that connects electronic-structure force inputs to KvN--Langevin canonical
phase-space state preparation and, within the same workflow, to both dynamical and
static molecular-dynamics observable readouts.

\begin{acknowledgments}
The author thanks Naoki Mitsumoto and Hidehiko Hiramatsu of DENSO CORPORATION, as well as the members of Quemix Inc., for valuable contributions and discussions. This work was supported by the Center of Innovation for Sustainable Quantum AI, JST Grant Number JPMJPF2221. The computation was performed using the facilities of the Supercomputer Center, the Institute for Solid State Physics, the University of Tokyo (ISSPkyodo-SC-2026-Ea-0014), the TSUBAME4.0 supercomputer at the Institute of Science Tokyo, and the Supermicro ARS-111GL-DNHR-LCC and FUJITSU Server PRIMERGY CX2550 M7 (Miyabi) at Joint Center for Advanced High Performance Computing (JCAHPC).
\end{acknowledgments}

\appendix
%==============================================================================
\section{\label{app:cos-bias}Leading-order cos-filter temperature-bias correction}
%==============================================================================

Here we derive the leading-order kinetic-temperature bias caused by replacing ideal
Gaussian diffusion with the PITE/cosinefilter. The analysis is performed in the
harmonic approximation and in the continuous-grid limit. We focus on the stationary
momentum distribution generated by the friction and diffusion blocks, which is the
part relevant to the internal-temperature correction used in
Sec.~\ref{sec:framework}.

Let $\kappa$ be the Fourier-conjugate variable to the momentum $P$, and let
$\widetilde{\psi}(\kappa)$ denote the KvN amplitude after Fourier transformation in
the $P$ direction. Since the friction step contracts the momentum distribution by
$e^{-\gamma\Delta t}$, its action in $\kappa$ space is
\begin{equation}
  \widetilde{\psi}(\kappa)
  \longmapsto
  e^{-s/2}\,\widetilde{\psi}(y\kappa),
  \qquad
  s \equiv \gamma\Delta t,
  \qquad
  y \equiv e^{-s}.
  \label{eq:app-friction-kappa}
\end{equation}
The PITE/cosinefilter acts as
\begin{equation}
  \widetilde{\psi}(\kappa)
  \longmapsto
  \cos(\sigma_{H}\kappa)\,\widetilde{\psi}(\kappa).
  \label{eq:app-cos-kappa}
\end{equation}
Thus, up to normalization, one friction-diffusion step gives
\begin{equation}
  \widetilde{\psi}_{n+1}(\kappa)
  =
  \cos(\sigma_{H}\kappa)\,
  e^{-s/2}\,
  \widetilde{\psi}_{n}(y\kappa).
  \label{eq:app-one-step-kappa}
\end{equation}
In the stationary state, the shape of the amplitude is reproduced after each step.
This gives
\begin{equation}
  \widetilde{\psi}_{\mathrm{eq}}(\kappa)
  \propto
  \prod_{r=0}^{\infty}
  \cos(\sigma_{H}y^{r}\kappa).
  \label{eq:app-cos-product}
\end{equation}

Taking the logarithm of Eq.~(\ref{eq:app-cos-product}) and using
$\log\cos x=-x^{2}/2-x^{4}/12+O(x^{6})$, we obtain
\begin{equation}
  \log\widetilde{\psi}_{\mathrm{eq}}(\kappa)
  =
  -A\kappa^{2}
  -
  B\kappa^{4}
  +
  O(\kappa^{6}),
  \label{eq:app-log-shape}
\end{equation}
where
\begin{equation}
  A
  =
  \frac{\sigma_{H}^{2}}{2(1-y^{2})},
  \qquad
  B
  =
  \frac{\sigma_{H}^{4}}{12(1-y^{4})}.
  \label{eq:app-AB}
\end{equation}
Using the discrete-time fluctuation--dissipation condition in the main text,
Eq.~(\ref{eq:lt-fdt}),
\begin{equation}
  \sigma_{H}^{2}
  =
  2\mu T_{\mathrm{int}}(1-y^{2}),
  \label{eq:app-lt-fdt}
\end{equation}
the Gaussian coefficient becomes
\begin{equation}
  A
  =
  \mu T_{\mathrm{int}}.
  \label{eq:app-A}
\end{equation}
If the quartic and higher-order terms were absent, the stationary kinetic
temperature would therefore be $T_{\mathrm{int}}$. The leading deviation from this
ideal Gaussian limit is generated by the $B\kappa^{4}$ term.

We treat $B/A^{2}$ as a small parameter and approximate the stationary amplitude as
\begin{equation}
  \widetilde{\psi}_{\mathrm{eq}}(\kappa)
  \simeq
  \exp\!\big(-A\kappa^{2}-B\kappa^{4}\big).
\end{equation}
By Parseval's relation,
\begin{equation}
  \langle P^{2}\rangle
  =
  \frac{
  \int d\kappa\;
  |\partial_{\kappa}\widetilde{\psi}_{\mathrm{eq}}(\kappa)|^{2}
  }{
  \int d\kappa\;
  |\widetilde{\psi}_{\mathrm{eq}}(\kappa)|^{2}
  }.
  \label{eq:app-parseval}
\end{equation}
Evaluating the Gaussian integrals to first order in $B$ gives
\begin{equation}
  \langle P^{2}\rangle
  =
  A
  +
  \frac{3B}{2A}
  +
  O\!\left(\frac{B^{2}}{A^{3}}\right).
  \label{eq:app-p2}
\end{equation}
From Eqs.~(\ref{eq:app-AB}) and (\ref{eq:app-lt-fdt}), we find
\begin{equation}
  \frac{3B}{2A^{2}}
  =
  \frac{1-y^{2}}{2(1+y^{2})}
  =
  \frac{1}{2}\tanh s.
  \label{eq:app-bias-factor}
\end{equation}
The kinetic temperature
$T_{\mathrm{kin}}\equiv\langle P^{2}\rangle/\mu$ therefore satisfies
\begin{equation}
  \frac{T_{\mathrm{kin}}-T_{\mathrm{int}}}{T_{\mathrm{int}}}
  =
  \frac{1}{2}\tanh(\gamma\Delta t)
  +
  O\!\big((\gamma\Delta t)^{2}\big).
  \label{eq:app-temp-bias}
\end{equation}
This result is Eq.~(\ref{eq:temperature-bias}) in the main text. The leading term
comes from the $x^{4}$ contribution in the expansion of $\log\cos x$ and is absent
for an ideal Gaussian heat kernel.

To match the kinetic temperature to the target physical temperature
$T_{\mathrm{phys}}$ at leading order, we choose the internal temperature supplied to
the circuit as
\begin{equation}
  T_{\mathrm{int}}
  =
  \frac{T_{\mathrm{phys}}}
  {1+\tfrac{1}{2}\tanh(\gamma\Delta t)}.
  \label{eq:app-Tint}
\end{equation}
This correction cancels the $O(\gamma\Delta t)$ temperature shift induced by the cos
filter. Residual deviations can still arise from higher-order terms in
$\gamma\Delta t$, finite-grid effects, anharmonicity of the potential, and finite
thermalization time. Figure~\ref{fig:canonical}(h) shows that, for the H$_2$
parameters used in this work, the corrected circuit yields a momentum distribution
near the target temperature.

%==============================================================================
\section{Reference protocols for the VDOS and TST readouts}
\label{app:aimd-postprocess}
\label{app:tst-details}
%==============================================================================

We first describe the AIMD reference used in Fig.~\ref{fig:vdos}(a). The reference
trajectories are generated by Hamiltonian dynamics on the same FCI/STO-3G
potential-energy surface as that used in the KvN calculation. Initial conditions are
sampled from the canonical distribution, and the internuclear distance
$R_{r}(t_{n})$ is recorded for each trajectory $r$. We define the centered coordinate
\begin{equation}
  Q_{r}(t_{n})
  =
  R_{r}(t_{n})-\langle R\rangle_{\mathrm{eq}},
\end{equation}
and estimate the trajectory-averaged autocorrelation function as
\begin{equation}
  C^{\mathrm{AIMD}}_{QQ}(t_{n})
  =
  \frac{1}{N_{\mathrm{traj}}}
  \sum_{r=1}^{N_{\mathrm{traj}}}
  Q_{r}(t_{n})Q_{r}(0).
  \label{eq:aimd-corr}
\end{equation}
In the continuous limit, when the same Born--Oppenheimer potential, canonical
distribution, and Hamiltonian dynamics are used, this AIMD correlation function gives
the same $R$-VDOS as the KvN $Q$-correlation defined in
Eq.~(\ref{eq:vdos-correlation}).

Because the correlation function is finite in time, we apply a window function
$w_{n}$ and compute
\begin{equation}
  S_{\mathrm{AIMD}}(\omega)
  =
  \left|
  \sum_{n=0}^{N_{t}-1}
  w_{n}
  C^{\mathrm{AIMD}}_{QQ}(t_{n})
  e^{i\omega t_{n}}
  \Delta t
  \right|^{2}.
  \label{eq:aimd-spectrum}
\end{equation}
For comparison with QPE, this AIMD spectrum is then matched to the same finite time
window and frequency binning. Let $\tau_{\mathrm{QPE}}$ be the QPE time interval,
$m$ the number of phase ancilla qubits, and $\omega_{\mathrm{shift}}$ the frequency
shift. The QPE display-window width, bin width, and bin centers are
\begin{equation}
  \Omega_{\mathrm{QPE}}
  =
  \frac{2\pi}{\tau_{\mathrm{QPE}}},
  \qquad
  \Delta\omega
  =
  \frac{\Omega_{\mathrm{QPE}}}{2^{m}},
  \qquad
  \omega_{j}
  =
  \omega_{\mathrm{shift}}+j\Delta\omega .
  \label{eq:qpe-bins}
\end{equation}
Frequency components separated by $\Omega_{\mathrm{QPE}}$ are folded into the same
display window. We therefore define the wrapped spectrum
\begin{equation}
  S_{\mathrm{wrap}}(\omega)
  =
  \sum_{\ell\in\mathbb{Z}}
  S_{\mathrm{AIMD}}(\omega+\ell\Omega_{\mathrm{QPE}})
  \label{eq:aimd-wrap}
\end{equation}
for
$\omega\in[\omega_{\mathrm{shift}},
\omega_{\mathrm{shift}}+\Omega_{\mathrm{QPE}})$. The AIMD reference value assigned to
the $j$th QPE bin is
\begin{equation}
  \bar{S}^{\mathrm{AIMD}}_{j}
  =
  \mathcal{N}
  \int_{\omega_{\mathrm{shift}}}^{\omega_{\mathrm{shift}}+\Omega_{\mathrm{QPE}}}
  d\omega\;
  F_{m}\!\big[(\omega-\omega_{j})\tau_{\mathrm{QPE}}\big]\,
  S_{\mathrm{wrap}}(\omega),
  \label{eq:aimd-qpe-bin}
\end{equation}
where $\mathcal{N}$ is a normalization constant used to match either the maximum
value or the total spectral weight. The finite-time QPE kernel is
\begin{equation}
  F_{m}(\theta)
  =
  \frac{1}{2^{m}}
  \left[
  \frac{\sin(2^{m}\theta/2)}{\sin(\theta/2)}
  \right]^{2}.
  \label{eq:fejer-kernel}
\end{equation}
The binned reference $\bar{S}^{\mathrm{AIMD}}_{j}$ in
Eq.~(\ref{eq:aimd-qpe-bin}) is the AIMD spectrum plotted against the KvN--QPE output
in Fig.~\ref{fig:vdos}(a).

We next summarize the grid estimator used for the TST readout in
Sec.~\ref{sec:tst}. The canonical probability assigned to grid point
$(R_{j},P_{\ell})$ is
\begin{equation}
  \begin{gathered}
  p_{j\ell}(T)
  =
  \frac{1}{Z}
  \exp\!\left[
  -\frac{1}{T}
  \left(
  \frac{P_{\ell}^{2}}{2\mu}+V(R_{j})
  \right)
  \right]
  \Delta R\,\Delta P,\\
  \sum_{j,\ell}p_{j\ell}(T)=1 .
  \end{gathered}
  \label{eq:app-canonical-grid}
\end{equation}
The analytically encoded canonical KvN state is taken to have zero phase,
$\psi_{j\ell}=\sqrt{p_{j\ell}}$. The positive-direction flux and the reactant
population are evaluated as
\begin{align}
  \Phi^{\ddagger}_{+}(T)
  &=
  \sum_{j,\ell}
  \delta_{\sigma}(R_{j}-R^{\ddagger})
  \Theta(P_{\ell})
  \frac{P_{\ell}}{\mu}
  p_{j\ell}(T),
  \label{eq:app-tst-flux-grid}\\
  P_{\mathrm{R}}(T)
  &=
  \sum_{R_{j}<R^{\ddagger},\,\ell}
  p_{j\ell}(T).
  \label{eq:app-tst-pop-grid}
\end{align}
Here $\delta_{\sigma}$ is a smoothed delta function representing the dividing surface
$R=R^{\ddagger}$ on the grid, and $\Theta$ is the Heaviside step function. The TST
rate reported in the main text is
$k_{\mathrm{TST}}=\Phi^{\ddagger}_{+}/P_{\mathrm{R}}$.

The green points in Fig.~\ref{fig:tst-rate} are obtained by evaluating
Eqs.~(\ref{eq:app-tst-flux-grid}) and (\ref{eq:app-tst-pop-grid}) from this
analytically encoded canonical KvN state. The right column of
Fig.~\ref{fig:phase-space} visualizes the same canonical density.

For the particle-based crossing reference, canonical initial conditions are generated
by Langevin dynamics or Monte Carlo. Hamiltonian trajectories are then propagated for
a finite time $t_{\mathrm{sim}}$, and positive crossings of the dividing surface
$R=R^{\ddagger}$ are counted. If $N_{\mathrm{cross}}$ crossings are observed from
$N_{\mathrm{traj}}$ trajectories, the finite-time crossing estimate is
\begin{equation}
  k_{\mathrm{cross}}
  =
  \frac{N_{\mathrm{cross}}}{N_{\mathrm{traj}}t_{\mathrm{sim}}}.
  \label{eq:app-crossing-rate}
\end{equation}
The corresponding smallest detectable value is estimated as
\begin{equation}
  k_{\mathrm{min}}
  \simeq
  \frac{N_{\mathrm{min}}}{N_{\mathrm{traj}}t_{\mathrm{sim}}},
  \label{eq:app-md-limit}
\end{equation}
where $N_{\mathrm{min}}$ is the minimum number of crossings required for detection.
This reference is not a count of reactions in Langevin trajectories. It is a
finite-time NVE crossing reference initialized from the canonical distribution.

\clearpage
\bibliographystyle{apsrev4-2}
\bibliography{refs}

\begin{thebibliography}{27}%
\makeatletter
\providecommand \@ifxundefined [1]{%
 \@ifx{#1\undefined}
}%
\providecommand \@ifnum [1]{%
 \ifnum #1\expandafter \@firstoftwo
 \else \expandafter \@secondoftwo
 \fi
}%
\providecommand \@ifx [1]{%
 \ifx #1\expandafter \@firstoftwo
 \else \expandafter \@secondoftwo
 \fi
}%
\providecommand \natexlab [1]{#1}%
\providecommand \enquote  [1]{``#1''}%
\providecommand \bibnamefont  [1]{#1}%
\providecommand \bibfnamefont [1]{#1}%
\providecommand \citenamefont [1]{#1}%
\providecommand \href@noop [0]{\@secondoftwo}%
\providecommand \href [0]{\begingroup \@sanitize@url \@href}%
\providecommand \@href[1]{\@@startlink{#1}\@@href}%
\providecommand \@@href[1]{\endgroup#1\@@endlink}%
\providecommand \@sanitize@url [0]{\catcode `\\12\catcode `\$12\catcode `\&12\catcode `\#12\catcode `\^12\catcode `\_12\catcode `\%12\relax}%
\providecommand \@@startlink[1]{}%
\providecommand \@@endlink[0]{}%
\providecommand \url  [0]{\begingroup\@sanitize@url \@url }%
\providecommand \@url [1]{\endgroup\@href {#1}{\urlprefix }}%
\providecommand \urlprefix  [0]{URL }%
\providecommand \Eprint [0]{\href }%
\providecommand \doibase [0]{https://doi.org/}%
\providecommand \selectlanguage [0]{\@gobble}%
\providecommand \bibinfo  [0]{\@secondoftwo}%
\providecommand \bibfield  [0]{\@secondoftwo}%
\providecommand \translation [1]{[#1]}%
\providecommand \BibitemOpen [0]{}%
\providecommand \bibitemStop [0]{}%
\providecommand \bibitemNoStop [0]{.\EOS\space}%
\providecommand \EOS [0]{\spacefactor3000\relax}%
\providecommand \BibitemShut  [1]{\csname bibitem#1\endcsname}%
\let\auto@bib@innerbib\@empty
%</preamble>
\bibitem [{\citenamefont {{Google Quantum AI and Collaborators}}(2025)}]{GoogleQAI2025Nature}%
  \BibitemOpen
  \bibfield  {author} {\bibinfo {author} {\bibnamefont {{Google Quantum AI and Collaborators}}},\ }\href {https://doi.org/10.1038/s41586-024-08449-y} {\bibfield  {journal} {\bibinfo  {journal} {Nature}\ }\textbf {\bibinfo {volume} {638}},\ \bibinfo {pages} {920} (\bibinfo {year} {2025})}\BibitemShut {NoStop}%
\bibitem [{\citenamefont {Bluvstein}\ \emph {et~al.}(2024)\citenamefont {Bluvstein} \emph {et~al.}}]{Bluvstein2024Nature}%
  \BibitemOpen
  \bibfield  {author} {\bibinfo {author} {\bibfnamefont {D.}~\bibnamefont {Bluvstein}} \emph {et~al.},\ }\href {https://doi.org/10.1038/s41586-023-06927-3} {\bibfield  {journal} {\bibinfo  {journal} {Nature}\ }\textbf {\bibinfo {volume} {626}},\ \bibinfo {pages} {58} (\bibinfo {year} {2024})}\BibitemShut {NoStop}%
\bibitem [{\citenamefont {Sales~Rodriguez}\ \emph {et~al.}(2025)\citenamefont {Sales~Rodriguez} \emph {et~al.}}]{Rodriguez2025Nature}%
  \BibitemOpen
  \bibfield  {author} {\bibinfo {author} {\bibfnamefont {P.}~\bibnamefont {Sales~Rodriguez}} \emph {et~al.},\ }\href {https://doi.org/10.1038/s41586-025-09367-3} {\bibfield  {journal} {\bibinfo  {journal} {Nature}\ }\textbf {\bibinfo {volume} {645}},\ \bibinfo {pages} {620} (\bibinfo {year} {2025})}\BibitemShut {NoStop}%
\bibitem [{\citenamefont {Ransford}\ \emph {et~al.}(2025)\citenamefont {Ransford} \emph {et~al.}}]{Ransford2025Helios}%
  \BibitemOpen
  \bibfield  {author} {\bibinfo {author} {\bibfnamefont {A.}~\bibnamefont {Ransford}} \emph {et~al.},\ }\href@noop {} {\bibfield  {journal} {\bibinfo  {journal} {arXiv preprint}\ }\textbf {\bibinfo {volume} {arXiv:2511.05465}} (\bibinfo {year} {2025})},\ \Eprint {https://arxiv.org/abs/2511.05465} {arXiv:2511.05465 [quant-ph]} \BibitemShut {NoStop}%
\bibitem [{\citenamefont {Montanaro}(2016)}]{Montanaro2016npjQI}%
  \BibitemOpen
  \bibfield  {author} {\bibinfo {author} {\bibfnamefont {A.}~\bibnamefont {Montanaro}},\ }\href {https://doi.org/10.1038/npjqi.2015.23} {\bibfield  {journal} {\bibinfo  {journal} {npj Quantum Inf.}\ }\textbf {\bibinfo {volume} {2}},\ \bibinfo {pages} {15023} (\bibinfo {year} {2016})}\BibitemShut {NoStop}%
\bibitem [{\citenamefont {McArdle}\ \emph {et~al.}(2020)\citenamefont {McArdle}, \citenamefont {Endo}, \citenamefont {Aspuru-Guzik}, \citenamefont {Benjamin},\ and\ \citenamefont {Yuan}}]{McArdle2020}%
  \BibitemOpen
  \bibfield  {author} {\bibinfo {author} {\bibfnamefont {S.}~\bibnamefont {McArdle}}, \bibinfo {author} {\bibfnamefont {S.}~\bibnamefont {Endo}}, \bibinfo {author} {\bibfnamefont {A.}~\bibnamefont {Aspuru-Guzik}}, \bibinfo {author} {\bibfnamefont {S.~C.}\ \bibnamefont {Benjamin}},\ and\ \bibinfo {author} {\bibfnamefont {X.}~\bibnamefont {Yuan}},\ }\href {https://doi.org/10.1103/RevModPhys.92.015003} {\bibfield  {journal} {\bibinfo  {journal} {Rev. Mod. Phys.}\ }\textbf {\bibinfo {volume} {92}},\ \bibinfo {pages} {015003} (\bibinfo {year} {2020})}\BibitemShut {NoStop}%
\bibitem [{\citenamefont {Aspuru-Guzik}\ \emph {et~al.}(2005)\citenamefont {Aspuru-Guzik}, \citenamefont {Dutoi}, \citenamefont {Love},\ and\ \citenamefont {Head-Gordon}}]{AspuruGuzik2005}%
  \BibitemOpen
  \bibfield  {author} {\bibinfo {author} {\bibfnamefont {A.}~\bibnamefont {Aspuru-Guzik}}, \bibinfo {author} {\bibfnamefont {A.~D.}\ \bibnamefont {Dutoi}}, \bibinfo {author} {\bibfnamefont {P.~J.}\ \bibnamefont {Love}},\ and\ \bibinfo {author} {\bibfnamefont {M.}~\bibnamefont {Head-Gordon}},\ }\href {https://doi.org/10.1126/science.1113479} {\bibfield  {journal} {\bibinfo  {journal} {Science}\ }\textbf {\bibinfo {volume} {309}},\ \bibinfo {pages} {1704} (\bibinfo {year} {2005})}\BibitemShut {NoStop}%
\bibitem [{\citenamefont {Car}\ and\ \citenamefont {Parrinello}(1985)}]{Car1985}%
  \BibitemOpen
  \bibfield  {author} {\bibinfo {author} {\bibfnamefont {R.}~\bibnamefont {Car}}\ and\ \bibinfo {author} {\bibfnamefont {M.}~\bibnamefont {Parrinello}},\ }\href {https://doi.org/10.1103/PhysRevLett.55.2471} {\bibfield  {journal} {\bibinfo  {journal} {Phys. Rev. Lett.}\ }\textbf {\bibinfo {volume} {55}},\ \bibinfo {pages} {2471} (\bibinfo {year} {1985})}\BibitemShut {NoStop}%
\bibitem [{\citenamefont {Marx}\ and\ \citenamefont {Hutter}(2009)}]{MarxHutter2009}%
  \BibitemOpen
  \bibfield  {author} {\bibinfo {author} {\bibfnamefont {D.}~\bibnamefont {Marx}}\ and\ \bibinfo {author} {\bibfnamefont {J.}~\bibnamefont {Hutter}},\ }\href {https://doi.org/10.1017/CBO9780511609633} {\emph {\bibinfo {title} {Ab Initio Molecular Dynamics: Basic Theory and Advanced Methods}}}\ (\bibinfo  {publisher} {Cambridge University Press},\ \bibinfo {address} {Cambridge},\ \bibinfo {year} {2009})\BibitemShut {NoStop}%
\bibitem [{\citenamefont {Tuckerman}(2023)}]{Tuckerman2010}%
  \BibitemOpen
  \bibfield  {author} {\bibinfo {author} {\bibfnamefont {M.~E.}\ \bibnamefont {Tuckerman}},\ }\href {https://doi.org/10.1093/oso/9780198825562.001.0001} {\emph {\bibinfo {title} {Statistical Mechanics: Theory and Molecular Simulation}}}\ (\bibinfo  {publisher} {Oxford University Press},\ \bibinfo {year} {2023})\BibitemShut {NoStop}%
\bibitem [{\citenamefont {Fedorov}\ \emph {et~al.}(2021)\citenamefont {Fedorov}, \citenamefont {Otten}, \citenamefont {Gray},\ and\ \citenamefont {Alexeev}}]{Fedorov2021JCP}%
  \BibitemOpen
  \bibfield  {author} {\bibinfo {author} {\bibfnamefont {D.~A.}\ \bibnamefont {Fedorov}}, \bibinfo {author} {\bibfnamefont {M.~J.}\ \bibnamefont {Otten}}, \bibinfo {author} {\bibfnamefont {S.~K.}\ \bibnamefont {Gray}},\ and\ \bibinfo {author} {\bibfnamefont {Y.}~\bibnamefont {Alexeev}},\ }\href {https://doi.org/10.1063/5.0046930} {\bibfield  {journal} {\bibinfo  {journal} {J. Chem. Phys.}\ }\textbf {\bibinfo {volume} {154}},\ \bibinfo {pages} {164103} (\bibinfo {year} {2021})}\BibitemShut {NoStop}%
\bibitem [{\citenamefont {Dononelli}(2026)}]{Dononelli2026}%
  \BibitemOpen
  \bibfield  {author} {\bibinfo {author} {\bibfnamefont {W.}~\bibnamefont {Dononelli}},\ }\href@noop {} {\bibfield  {journal} {\bibinfo  {journal} {arXiv preprint}\ }\textbf {\bibinfo {volume} {arXiv:2602.02695}} (\bibinfo {year} {2026})},\ \Eprint {https://arxiv.org/abs/2602.02695} {arXiv:2602.02695 [quant-ph]} \BibitemShut {NoStop}%
\bibitem [{\citenamefont {Kassal}\ \emph {et~al.}(2008)\citenamefont {Kassal}, \citenamefont {Jordan}, \citenamefont {Love}, \citenamefont {Mohseni},\ and\ \citenamefont {Aspuru-Guzik}}]{Kassal2008}%
  \BibitemOpen
  \bibfield  {author} {\bibinfo {author} {\bibfnamefont {I.}~\bibnamefont {Kassal}}, \bibinfo {author} {\bibfnamefont {S.~P.}\ \bibnamefont {Jordan}}, \bibinfo {author} {\bibfnamefont {P.~J.}\ \bibnamefont {Love}}, \bibinfo {author} {\bibfnamefont {M.}~\bibnamefont {Mohseni}},\ and\ \bibinfo {author} {\bibfnamefont {A.}~\bibnamefont {Aspuru-Guzik}},\ }\href {https://doi.org/10.1073/pnas.0808245105} {\bibfield  {journal} {\bibinfo  {journal} {Proc. Natl. Acad. Sci. USA}\ }\textbf {\bibinfo {volume} {105}},\ \bibinfo {pages} {18681} (\bibinfo {year} {2008})}\BibitemShut {NoStop}%
\bibitem [{\citenamefont {da~Jornada}\ \emph {et~al.}(2025)\citenamefont {da~Jornada}, \citenamefont {Lostaglio}, \citenamefont {Pallister}, \citenamefont {{\c S}ahino{\u g}lu},\ and\ \citenamefont {Seetharam}}]{daJornada2025FTQC}%
  \BibitemOpen
  \bibfield  {author} {\bibinfo {author} {\bibfnamefont {F.~H.}\ \bibnamefont {da~Jornada}}, \bibinfo {author} {\bibfnamefont {M.}~\bibnamefont {Lostaglio}}, \bibinfo {author} {\bibfnamefont {S.}~\bibnamefont {Pallister}}, \bibinfo {author} {\bibfnamefont {B.}~\bibnamefont {{\c S}ahino{\u g}lu}},\ and\ \bibinfo {author} {\bibfnamefont {K.~I.}\ \bibnamefont {Seetharam}},\ }\href@noop {} {\bibfield  {journal} {\bibinfo  {journal} {arXiv preprint}\ }\textbf {\bibinfo {volume} {arXiv:2504.06348}} (\bibinfo {year} {2025})},\ \Eprint {https://arxiv.org/abs/2504.06348} {arXiv:2504.06348 [quant-ph]} \BibitemShut {NoStop}%
\bibitem [{\citenamefont {Pocrnic}\ \emph {et~al.}(2026)\citenamefont {Pocrnic}, \citenamefont {Loaiza}, \citenamefont {Arrazola}, \citenamefont {Wiebe},\ and\ \citenamefont {Motlagh}}]{Pocrnic2026preBO}%
  \BibitemOpen
  \bibfield  {author} {\bibinfo {author} {\bibfnamefont {M.}~\bibnamefont {Pocrnic}}, \bibinfo {author} {\bibfnamefont {I.}~\bibnamefont {Loaiza}}, \bibinfo {author} {\bibfnamefont {J.~M.}\ \bibnamefont {Arrazola}}, \bibinfo {author} {\bibfnamefont {N.}~\bibnamefont {Wiebe}},\ and\ \bibinfo {author} {\bibfnamefont {D.}~\bibnamefont {Motlagh}},\ }\href@noop {} {\bibfield  {journal} {\bibinfo  {journal} {arXiv preprint}\ }\textbf {\bibinfo {volume} {arXiv:2602.11272}} (\bibinfo {year} {2026})},\ \Eprint {https://arxiv.org/abs/2602.11272} {arXiv:2602.11272 [quant-ph]} \BibitemShut {NoStop}%
\bibitem [{\citenamefont {Eklund}\ \emph {et~al.}(2026)\citenamefont {Eklund}, \citenamefont {Tikku}, \citenamefont {Sinnott}, \citenamefont {Huggins}, \citenamefont {Low}, \citenamefont {Berry},\ and\ \citenamefont {Kassal}}]{Eklund2026endtoend}%
  \BibitemOpen
  \bibfield  {author} {\bibinfo {author} {\bibfnamefont {E.~C.}\ \bibnamefont {Eklund}}, \bibinfo {author} {\bibfnamefont {A.}~\bibnamefont {Tikku}}, \bibinfo {author} {\bibfnamefont {P.}~\bibnamefont {Sinnott}}, \bibinfo {author} {\bibfnamefont {W.~J.}\ \bibnamefont {Huggins}}, \bibinfo {author} {\bibfnamefont {G.~H.}\ \bibnamefont {Low}}, \bibinfo {author} {\bibfnamefont {D.~W.}\ \bibnamefont {Berry}},\ and\ \bibinfo {author} {\bibfnamefont {I.}~\bibnamefont {Kassal}},\ }\href@noop {} {\bibfield  {journal} {\bibinfo  {journal} {arXiv preprint}\ }\textbf {\bibinfo {volume} {arXiv:2603.19007}} (\bibinfo {year} {2026})},\ \Eprint {https://arxiv.org/abs/2603.19007} {arXiv:2603.19007 [quant-ph]} \BibitemShut {NoStop}%
\bibitem [{\citenamefont {Koopman}(1931)}]{Koopman1931}%
  \BibitemOpen
  \bibfield  {author} {\bibinfo {author} {\bibfnamefont {B.~O.}\ \bibnamefont {Koopman}},\ }\href {https://doi.org/10.1073/pnas.17.5.315} {\bibfield  {journal} {\bibinfo  {journal} {Proc. Natl. Acad. Sci. U.S.A.}\ }\textbf {\bibinfo {volume} {17}},\ \bibinfo {pages} {315} (\bibinfo {year} {1931})}\BibitemShut {NoStop}%
\bibitem [{\citenamefont {v.~Neumann}(1932)}]{vonNeumann1932}%
  \BibitemOpen
  \bibfield  {author} {\bibinfo {author} {\bibfnamefont {J.}~\bibnamefont {v.~Neumann}},\ }\href {http://www.jstor.org/stable/1968537} {\bibfield  {journal} {\bibinfo  {journal} {Annals of Mathematics}\ }\textbf {\bibinfo {volume} {33}},\ \bibinfo {pages} {587} (\bibinfo {year} {1932})}\BibitemShut {NoStop}%
\bibitem [{\citenamefont {Joseph}(2020)}]{Joseph2020PRR}%
  \BibitemOpen
  \bibfield  {author} {\bibinfo {author} {\bibfnamefont {I.}~\bibnamefont {Joseph}},\ }\href {https://doi.org/10.1103/PhysRevResearch.2.043102} {\bibfield  {journal} {\bibinfo  {journal} {Phys. Rev. Research}\ }\textbf {\bibinfo {volume} {2}},\ \bibinfo {pages} {043102} (\bibinfo {year} {2020})}\BibitemShut {NoStop}%
\bibitem [{\citenamefont {Watanabe}\ \emph {et~al.}(2026)\citenamefont {Watanabe}, \citenamefont {Nishi}, \citenamefont {Kosugi}, \citenamefont {Hidaka}, \citenamefont {Sakurai},\ and\ \citenamefont {Matsushita}}]{PaperI}%
  \BibitemOpen
  \bibfield  {author} {\bibinfo {author} {\bibfnamefont {M.}~\bibnamefont {Watanabe}}, \bibinfo {author} {\bibfnamefont {H.}~\bibnamefont {Nishi}}, \bibinfo {author} {\bibfnamefont {T.}~\bibnamefont {Kosugi}}, \bibinfo {author} {\bibfnamefont {S.}~\bibnamefont {Hidaka}}, \bibinfo {author} {\bibfnamefont {R.}~\bibnamefont {Sakurai}},\ and\ \bibinfo {author} {\bibfnamefont {Y.-i.}\ \bibnamefont {Matsushita}},\ }\href@noop {} {\bibfield  {journal} {\bibinfo  {journal} {arXiv preprint}\ } (\bibinfo {year} {2026})},\ \bibinfo {note} {submitted May 28, 2026; arXiv identifier to be added}\BibitemShut {NoStop}%
\bibitem [{\citenamefont {Risken}(1989)}]{Risken1989}%
  \BibitemOpen
  \bibfield  {author} {\bibinfo {author} {\bibfnamefont {H.}~\bibnamefont {Risken}},\ }\href {https://doi.org/10.1007/978-3-642-61544-3} {\emph {\bibinfo {title} {The Fokker--Planck Equation: Methods of Solution and Applications}}},\ \bibinfo {edition} {2nd}\ ed.\ (\bibinfo  {publisher} {Springer},\ \bibinfo {address} {Berlin, Heidelberg},\ \bibinfo {year} {1989})\BibitemShut {NoStop}%
\bibitem [{\citenamefont {Leimkuhler}\ and\ \citenamefont {Matthews}(2015)}]{LeimkuhlerMatthews}%
  \BibitemOpen
  \bibfield  {author} {\bibinfo {author} {\bibfnamefont {B.}~\bibnamefont {Leimkuhler}}\ and\ \bibinfo {author} {\bibfnamefont {C.}~\bibnamefont {Matthews}},\ }\href {https://doi.org/10.1007/978-3-319-16375-8} {\emph {\bibinfo {title} {Molecular Dynamics: With Deterministic and Stochastic Numerical Methods}}},\ \bibinfo {series} {Interdisciplinary Applied Mathematics}, Vol.~\bibinfo {volume} {39}\ (\bibinfo  {publisher} {Springer},\ \bibinfo {address} {Cham},\ \bibinfo {year} {2015})\ \bibinfo {note} {see Chapter 7 for Langevin integrator analysis.}\BibitemShut {Stop}%
\bibitem [{\citenamefont {Trotter}(1959)}]{Trotter1959}%
  \BibitemOpen
  \bibfield  {author} {\bibinfo {author} {\bibfnamefont {H.~F.}\ \bibnamefont {Trotter}},\ }\href {https://doi.org/10.1090/S0002-9939-1959-0108732-6} {\bibfield  {journal} {\bibinfo  {journal} {Proceedings of the American Mathematical Society}\ }\textbf {\bibinfo {volume} {10}},\ \bibinfo {pages} {545} (\bibinfo {year} {1959})}\BibitemShut {NoStop}%
\bibitem [{\citenamefont {Suzuki}(1976)}]{Suzuki1976}%
  \BibitemOpen
  \bibfield  {author} {\bibinfo {author} {\bibfnamefont {M.}~\bibnamefont {Suzuki}},\ }\href {https://doi.org/10.1007/BF01609348} {\bibfield  {journal} {\bibinfo  {journal} {Communications in Mathematical Physics}\ }\textbf {\bibinfo {volume} {51}},\ \bibinfo {pages} {183} (\bibinfo {year} {1976})}\BibitemShut {NoStop}%
\bibitem [{\citenamefont {Suzuki}(1990)}]{suzuki1990}%
  \BibitemOpen
  \bibfield  {author} {\bibinfo {author} {\bibfnamefont {M.}~\bibnamefont {Suzuki}},\ }\href {https://doi.org/10.1016/0375-9601(90)90962-N} {\bibfield  {journal} {\bibinfo  {journal} {Physics Letters A}\ }\textbf {\bibinfo {volume} {146}},\ \bibinfo {pages} {319} (\bibinfo {year} {1990})}\BibitemShut {NoStop}%
\bibitem [{\citenamefont {Bravyi}\ \emph {et~al.}(2017)\citenamefont {Bravyi}, \citenamefont {Gambetta}, \citenamefont {Mezzacapo},\ and\ \citenamefont {Temme}}]{Bravyi2017taper}%
  \BibitemOpen
  \bibfield  {author} {\bibinfo {author} {\bibfnamefont {S.}~\bibnamefont {Bravyi}}, \bibinfo {author} {\bibfnamefont {J.~M.}\ \bibnamefont {Gambetta}}, \bibinfo {author} {\bibfnamefont {A.}~\bibnamefont {Mezzacapo}},\ and\ \bibinfo {author} {\bibfnamefont {K.}~\bibnamefont {Temme}},\ }\href@noop {} {\bibfield  {journal} {\bibinfo  {journal} {arXiv preprint}\ }\textbf {\bibinfo {volume} {arXiv:1701.08213}} (\bibinfo {year} {2017})},\ \Eprint {https://arxiv.org/abs/1701.08213} {arXiv:1701.08213 [quant-ph]} \BibitemShut {NoStop}%
\bibitem [{\citenamefont {Huang}\ \emph {et~al.}(2025)\citenamefont {Huang}, \citenamefont {Boyd}, \citenamefont {Anselmetti}, \citenamefont {Degroote}, \citenamefont {Moll}, \citenamefont {Santagati}, \citenamefont {Streif}, \citenamefont {Ries}, \citenamefont {Marti-Dafcik}, \citenamefont {Jnane}, \citenamefont {Simon}, \citenamefont {Wiebe}, \citenamefont {Bromley},\ and\ \citenamefont {Koczor}}]{Huang2025alchemist}%
  \BibitemOpen
  \bibfield  {author} {\bibinfo {author} {\bibfnamefont {P.-W.}\ \bibnamefont {Huang}}, \bibinfo {author} {\bibfnamefont {G.}~\bibnamefont {Boyd}}, \bibinfo {author} {\bibfnamefont {G.-L.~R.}\ \bibnamefont {Anselmetti}}, \bibinfo {author} {\bibfnamefont {M.}~\bibnamefont {Degroote}}, \bibinfo {author} {\bibfnamefont {N.}~\bibnamefont {Moll}}, \bibinfo {author} {\bibfnamefont {R.}~\bibnamefont {Santagati}}, \bibinfo {author} {\bibfnamefont {M.}~\bibnamefont {Streif}}, \bibinfo {author} {\bibfnamefont {B.}~\bibnamefont {Ries}}, \bibinfo {author} {\bibfnamefont {D.}~\bibnamefont {Marti-Dafcik}}, \bibinfo {author} {\bibfnamefont {H.}~\bibnamefont {Jnane}}, \bibinfo {author} {\bibfnamefont {S.}~\bibnamefont {Simon}}, \bibinfo {author} {\bibfnamefont {N.}~\bibnamefont {Wiebe}}, \bibinfo {author} {\bibfnamefont {T.~R.}\ \bibnamefont {Bromley}},\ and\ \bibinfo {author} {\bibfnamefont {B.}~\bibnamefont {Koczor}},\ }\href@noop {} {\bibfield  {journal} {\bibinfo  {journal} {arXiv preprint}\ }\textbf {\bibinfo {volume}
  {arXiv:2508.16719}} (\bibinfo {year} {2025})},\ \Eprint {https://arxiv.org/abs/2508.16719} {arXiv:2508.16719 [quant-ph]} \BibitemShut {NoStop}%
\end{thebibliography}%

\end{document}